\documentclass[sigplan,10pt]{acmart}
\usepackage{lingmacros}
\usepackage{tree-dvips}
\usepackage{url}
\usepackage{subfigure}

\usepackage{tikz}
\usepackage{listings}
\usepackage{colortbl}
\usepackage{graphicx}
\usepackage{multicol}
\usepackage{tabularx}
\usepackage{multirow}
\usepackage{enumitem}
\usepackage{amsmath}
\usepackage{longtable}
\usepackage{stfloats}
\usepackage{pgfplots}
\usetikzlibrary{pgfplots.groupplots}

\pgfplotsset{
    every non boxed x axis/.style={} 
}

\usetikzlibrary{arrows, chains, positioning, shapes.geometric, shapes.symbols}
\usetikzlibrary{fadings,calc,patterns}

\pgfplotsset{compat=1.17}

\acmConference{Jan}{2023}{arxiv}
\startPage{1}

\setcopyright{none}
\definecolor{dkred}{rgb}{0.6,0,0}
\definecolor{dkgreen}{rgb}{0,0.6,0}
\definecolor{dkblue}{rgb}{0.05,0.05,0.43}
\definecolor{mauve}{rgb}{0.6,0,0.6}

\definecolor{dkred}{rgb}{0.6,0,0}
\definecolor{dkgreen}{rgb}{0,0.6,0}
\definecolor{dkblue}{rgb}{0.05,0.05,0.43}
\definecolor{mauve}{rgb}{0.6,0,0.6}
\definecolor{ltblue}{rgb}{0.93,0.95,0.99}
\definecolor{ltorange}{rgb}{0.99,0.95,0.93}

\definecolor{blue1}{rgb}{0.5,0.65,0.83}
\definecolor{blue2}{rgb}{0.7,0.79,0.9}
\definecolor{red1}{rgb}{0.85,0.2,0.2}
\definecolor{red2}{rgb}{0.97,0.5,0.5}
\definecolor{green1}{rgb}{0.1,0.6,0.1}
\definecolor{green2}{rgb}{0.5,0.99,0.5}
\definecolor{dkblue1}{rgb}{0.07,0.04,0.54}
\definecolor{dkblue2}{rgb}{0.59,0.59,0.99}
\definecolor{orange1}{rgb}{1.0,0.66,0.16}
\definecolor{orange2}{rgb}{1.0,0.9,0.5}
\definecolor{gray1}{rgb}{0.92,0.92,0.92}
\definecolor{gray2}{rgb}{0.97,0.97,0.97}
\definecolor{gray3}{rgb}{0.07,0.07,0.07}
\definecolor{gray4}{rgb}{0.47,0.47,0.47}

\definecolor{Color1}{rgb}{0.95,0.95,0.95}
\definecolor{Color2}{rgb}{0.9,0.9,0.9}
\definecolor{HeadColor}{rgb}{0.5,0.5,0.5}

\tikzset{
empty/.style     = {font=\fontsize{6}{2.5}\selectfont},
empty1/.style     = {font=\fontsize{8}{8}\selectfont},
arrow/.style     = {font=\fontsize{4}{2.5}\selectfont},
base/.style      = {draw=black, inner sep=3pt, minimum width=1.5cm, minimum height=0.43cm, font=\fontsize{6}{2.5}\selectfont},
myarrows/.style  = {-stealth, thick, dkblue, text=black, font=\fontsize{3}{2}\selectfont},
style1/.style    = {rectangle, rounded corners, base,  shade, shading=axis, left color=blue2!10!white, right color=blue1, shading angle=0},
style2/.style    = {rectangle, rounded corners, base, minimum width=6cm,  shade, shading=axis, left color=blue1!20, right color=blue2!30, shading angle=0},
style3/.style    = {ellipse, base, inner sep=0pt, minimum width=2cm, fill=blue2, font=\fontsize{3.5}{2}\selectfont},
style4/.style    = {rectangle, rounded corners, base, shade, shading=axis, left color=gray1, right color=gray2, shading angle=0},
style5/.style    = {rectangle, rounded corners, base, shade, shading=axis, left color=gray3, right color=gray4, shading angle=0},
style6/.style    = {diamond, rounded corners, base, shade, shading=axis, left color=orange1!80!white, right color=orange2!10!white, shading angle=0},
style7/.style    = {cylinder, base, shape border rotate=90, draw, minimum height=2cm, cylinder uses custom fill, cylinder body fill = orange1, cylinder end fill = orange2},
line/.style      = {draw, myarrows},
cir/.style      = {circle, draw=black, inner sep=0pt, minimum width=0.5cm, minimum height=0.5cm},
smallcir/.style      = {cir, top color=orange, bottom color=red!5, minimum width=4, minimum height=4},
smallcir2/.style      = {cir, right color=dkblue!40, left color=orange1!80, minimum width=4, minimum height=4},
smallcir3/.style      = {cir, right color=dkblue!60, left color=orange1!60, minimum width=4, minimum height=4},
smallcir4/.style      = {cir, right color=dkblue!80, left color=orange1!40, minimum width=4, minimum height=4},
smallcir5/.style      = {cir, right color=dkblue, left color=orange1!20, minimum width=4, minimum height=4},
myarrows/.style  = {-stealth, thick, black, text=black, font=\fontsize{3}{2}\selectfont},
grid/.style      = {base, minimum width=0.5, minimum height=0.5, font=\fontsize{5}{2.5}\selectfont},
}

\newcommand{\gear}[7]{
  node[font=\fontsize{4}{6}\selectfont, text width=0.5cm, align=center] (#6) {#7}
  (0:#2)
  \foreach \i [evaluate=\i as \n using {\i-1)*360/#1}] in {1,...,#1}{
    arc (\n:\n+#4:#2) {[rounded corners=1.5pt] -- (\n+#4+#5:#3)
    arc (\n+#4+#5:\n+360/#1-#5:#3)} --  (\n+360/#1:#2)
  }
}

\tikzstyle{every node}=[font=\tiny]

\newcolumntype{L}[1]{>{\raggedright\let\newline\\\arraybackslash\hspace{0pt}}m{#1}}
\newcolumntype{R}[1]{>{\raggedleft\let\newline\\\arraybackslash\hspace{0pt}}m{#1}}
\newcolumntype{C}[1]{>{\centering\let\newline\\\arraybackslash\hspace{0pt}}m{#1}}

\title[OpenMP Advisor]{OpenMP Advisor}

\author{Alok Mishra}
\affiliation{
  \department{Computer Science}
  \institution{Stony Brook University}
  \city{Stony Brook}
  \state{NY}
  \country{USA}
}
\email{almishra@cs.stonybrook.edu}

\author{Abid M. Malik}
\affiliation{
  \department{Computer Science Initiative}
  \institution{Brookhaven National Laboratory}
  \city{Upton}
  \state{NY}
  \postcode{11973}
  \country{USA}
}
\email{amalik@bnl.gov}

\author{Meifeng Lin}
\affiliation{
  \department{Computer Science Initiative}
  \institution{Brookhaven National Laboratory}
  \city{Upton}
  \state{New York}
  \country{USA}
  \postcode{11973}
}
\email{mlin@bnl.gov}

\author{Barbara Chapman}
\affiliation{
  \department{Computer Science}
  \institution{Stony Brook University}
  \city{Stony Brook}
  \state{New York}
  \country{USA}
  \postcode{11794}
}
\email{barbara.chapman@stonybrook.edu}

\begin{document}
\begin{abstract}
With the increasing diversity of heterogeneous architecture in the HPC industry, porting a legacy application to run on different architectures is a tough challenge.
In this paper, we present OpenMP Advisor, a first of its kind compiler tool that enables code offloading to a GPU with OpenMP using Machine Learning.
Although the tool is currently limited to GPUs, it can be extended to support other OpenMP-capable devices.
The tool has two modes: Training mode and Prediction mode. 
The training mode must be executed on the target hardware. 
It takes benchmark codes as input, generates and executes every variant of the code that could possibly run on the target device, and then collects data from all of the executed codes to train an ML-based cost model for use in prediction mode.
However, in prediction mode the tool does not need any interaction with the target device.
It accepts a C code as input and returns the best code variant that can be used to offload the code to the specified device.
The tool can determine the kernels that are best suited for offloading by predicting their runtime using a machine learning-based cost model.
The main objective behind this tool is to maintain the portability aspect of OpenMP.
Using our Advisor, we were able to generate code of multiple applications for seven different architectures, and correctly predict the top ten best variants for each application on every architecture.
Preliminary findings indicate that this tool can assist compiler developers and HPC application researchers in porting their legacy HPC codes to the upcoming heterogeneous computing environment.
\end{abstract}
\maketitle

\section{Introduction}
Over two decades have passed since what has been referred to as ``\textit{the death of Moore's law}''~\cite{tuomi2002lives} and the transition to multi-core processors from ever-faster single chips.
However, given the increasing number of transistors on a chip, one could argue that Moore's law has not yet been completely broken; rather, it is no longer possible to run these transistors at ever-increasing speeds.
This is a result of \textit{Dennard scaling}~\cite{dennard1974design}, a related effect that occurred at the same time as the reduction in transistor size.
The main outcome of Dennard scaling was that power density remained constant as transistor size decreased.
Dennard had found that because voltage and current scale proportionally with feature size and power scales proportionally with area, performance per watt doubles roughly every two years.
Consequently, what went wrong was not the ability to etch smaller transistors, but rather the capacity to reduce the voltage and current they require to function properly.

Developers in the HPC industry have started looking beyond Moore's law and  hardware developers have improved chip performance by configuring a growing number of compute cores. 
This rapid change in architecture hardware necessitates programming language updates and modification of application code to exploit the cores, e.g. by inserting pthreads or OpenMP~\cite{dagum1998openmp} constructs into the source code.
The HPC community was quick to embraced this multi-core processor technology.

In the last decade, General Purpose Graphics Processing Units (GPGPUs) have been attached to the multi-core processors on many HPC platforms in order to benefit from their ability to handle large amounts of data parallelism with low power consumption.
Initially intended for graphics operations, they are now being used extensively by HPC clusters and for general-purpose computing on graphics processing units.
Although there are some notable exceptions, the most recent TOP500 list~\cite{top500list} clearly demonstrates the trend toward heterogeneous HPC platforms, particularly those using GPUs. 
A significant portion of the systems are heterogeneous, with AMD or NVIDIA GPUs typically providing high performance per watt.

\subsection{The Challenge}
Many programmers are adapting their code to take advantage of GPUs. 
Unfortunately, optimizing the computational power of a GPU while minimizing overheads is a laborious process that may necessitate re-engineering large sections of code and data structures.
The development of code for systems with extreme heterogeneity and numerous devices will be much more challenging.
Therefore, it is essential to create tools that will relieve the application scientists of the burden of such development.

Despite the variety of programming models available, it is still quite challenging to optimize large scale applications consisting of tens-to-hundreds of thousands lines of code. 
Even when using a directive based programming model such as OpenMP, pragmatizing each kernel is a repetitive and complex task. 
OpenMP offers a variety of options for offloading a kernel to GPUs.
However, the application scientist must still figure out all the intricate GPU configurations.
To demonstrate the complexity of porting a kernel to emerging exascale hardwares, we use a kernel from the Lattice Quantum Chromodynamics (LQCD)~\cite{refId0} application, which is a computer-friendly numerical framework for QCD. 
One of LQCD's key computational kernels is the Wilson Dslash operator~\cite{lin2016optimization}, which is essentially a finite difference operator, to describe the interaction of quarks with gluons.

The Wilson Dslash operator, $D$, in four space-time dimensions is defined by Equation~\ref{eq:dwf}.
\begin{equation}\label{eq:dwf}
\begin{aligned}
D_{\alpha \beta}^{ij}(x, y) =  \sum_{\mu=1}^{4}[&((1-\gamma_{\mu}))_{\alpha\beta}U_{\mu}^{ij}(x)\delta_{x+\hat{\mu},y} \\ 
& + (1+\gamma_{\mu})_{\alpha\beta}U_{\mu}^{\dagger^{ij}} (x+ \hat{\mu})\delta_{x-\hat{\mu},y}) ]
\end{aligned}
\end{equation}

Here $x$ and $y$ are the coordinates of the lattice sites, $\alpha, \beta$ are spin indices, and $i, j$ are color indices.
$U_{\mu} (x)$ is the gluon field variable and is an $SU(3)$ matrix. $\gamma_\mu$'s are $4\times4$ Dirac matrices that are fixed. 
The complex fermion fields are represented as one-dimensional arrays of size $L_X \times L_Y \times L_Z \times L_T \times SPINS \times COLORS \times 2$ for the unpreconditioned Dirac operator, where $L_X ,L_Y ,L_Z$ and $L_T$ are the numbers of lattice sites in the $x, y, z$ and $t$ directions, respectively. 
The spin and color degrees of freedom, which are commonly 4 and 3, are denoted by the variables $SPINS$ and $COLORS$.

\begin{figure}[t!]
\begin{lstlisting}[caption={Loops of Wilson Dslash Operator}, label=lst:advisor:dslash,frame=tlrb, captionpos=b, belowskip=-1\baselineskip]
<@\color{mauve}{\#pragma omp target}@>
  for(int i=0; i<N_i; i++) {
    for(int j=0; j<N_j; j++) {
      for(int k=0; k<N_k; k++) {
        for(int l=0; l<N_l; l++) {
          /* ... COMPUTE ... */
        }
      }
    }
  }
\end{lstlisting}
\end{figure}
When we express Equation~\ref{eq:dwf} in C++, it has four nested \texttt{for} loops iterating over $L_T, L_Z, L_Y,$ and $L_X$, as shown in Code~\ref{lst:advisor:dslash}.
When we keep the values of $L_T, L_Z, L_Y,$ and $L_X$ at $16$ each, the COMPUTE section of the code has over 5 million variable definitions, 1.2 billion variable references, over 150 million addition/subtraction, 163 million multiplication, and so on.
Additionally, this function is called several times throughout the LQCD application. 

It is a herculean task for an application scientist to understand the physics, transform it into computer program, analyze the offloadable kernel, and then consider how to parallelize it to execute efficiently on an HPC cluster.
To get the best performance out of a GPU, an application scientist needs a thorough understanding of the underlying architecture, algorithm, and interface programming model. 
Alternately, they could test out various GPU transformations until they find the most effective one. 
However, none of these strategies is very efficient.

\subsection{Our Contribution}
This paper presents \textbf{OpenMP Advisor}, a first-of-its-kind compiler tool that advises application scientists on various OpenMP code offloading strategies.
This tool performs the following tasks to successfully address the challenges of effectively transforming an OpenMP code:
\begin{enumerate}[nosep,leftmargin=*]
\item detect potentially offloadable kernel;
\item identify data access and modification in kernel;
\item recommend several potential OpenMP variants for offloading that kernel to the GPU;
\item evaluate the profitability of each kernel variant via an adaptive cost model;
\item insert pertinent OpenMP directives to perform offloading.
\end{enumerate}

Although the tool is currently limited to GPUs, it can be extended to support other OpenMP-capable devices.
In the rest of the paper, we first discuss state of the art work that is related to and precedes our work in Section~\ref{sec:relatedwork}.
Then we define our OpenMP Advisor in Section~\ref{sec:ompadvisor}, and its implementation in Section~\ref{sec:impl}.
Section~\ref{sec:experiments} covers the experiments carried out in this paper and the analysis of the result.
Finally, we conclude our work with discussions of future plans in Section~\ref{sec:conclusion}.
\section{Related Work}
\label{sec:relatedwork}

Many studies have looked into how to best manage GPU memory when data movement must be explicitly managed.
A fully automatic system for managing and optimizing CPU-GPU communication for CUDA programs is provided by Jablin \emph{et.al.}~\cite{jablin2011automatic}.
Gelado \emph{et.al.}~\cite{gelado2010asymmetric} present a heterogeneous computing programming model to simplify and optimize GPU data management.
OMPSan~\cite{barua2019ompsan} performs static analysis on explicit data transfers that have already been inserted into an OpenMP code.
However, these studies do not provide insight into how data must be transferred and how data reusability on GPU can be used for implicitly managed data between multiple kernels.
Mishra et.al.~\cite{mishra2020data} proposed a technique for statically identifying data used by each kernel and automatically recognizing data reuse opportunities across kernels.
In our work, we use this technique to manage data between the CPU and the GPU.

Recently, Roy \emph{et.al.}~\cite{roy2021bliss} developed BLISS, a novel approach for auto-tuning parallel applications without the need for instrumentation, domain-specific knowledge, or prior knowledge of the applications.
With the help of Bayesian optimization, this auto-tuner creates a collection of various lightweight models that can rival the output quality of a complex, heavyweight Machine Learning (ML) model.
Other autotuners like BOHB\cite{falkner2018bohb}, HiPerBOt~\cite{menon2020auto}, GPTune~\cite{liu2021gptune} and ytopt~\cite{wu2022autotuning} frequently use Bayesian optimization in this context.
However, due to their expensive evaluation functions, tuning large-scale applications remains a challenge.
Besides, the need to optimize HPC programs on heterogeneous hardware (such as GPUs) and software configurations prevents the widespread use of these auto-tuning techniques in the Exascale era.
HPC applications are getting extremely heterogeneous, complicated, and increasingly expensive to analyze.

There is a need for a tool like ours that assists application scientists to offload their code to GPUs.
Other related research on automatic GPU offloading by Mendon{\c{c}}a \emph{et.al.}~\cite{mendoncca2017dawncc} and Poesia \emph{et.al.}~\cite{poesia2017static} can benefit from our tool by including our technique of data optimization and cost model in their framework, further reducing the challenges of using GPUs for scientific computing.
However, developing a cost model is time-consuming, and practically all contemporary compilers, such as LLVM/Clang~\cite{llvmcostmodel}, adopt a straightforward ``one-size-fits-all'' cost function that does not perform optimally in the situation of extreme heterogeneous architecture.
With the use of a neural network model, COMPOFF~\cite{mishra2022compoff} offers a new portable cost model that statically estimates the cost of OpenMP offloading on different architectures.
We extend the techniques defined in COMPOFF to develop a portable static cost model to be used in our OpenMP Advisor tool.
\section{OpenMP Advisor}
\label{sec:ompadvisor}

We design and develop the OpenMP Advisor, a compiler tool which transforms an OpenMP code to effectively offload to a GPU.
This tool identifies OpenMP kernels, suggest a number of possible OpenMP variants for offloading that kernel to the GPU, and predict the cost of running each of these kernels separately using a static neural network-based compile time cost model.
The OpenMP Advisor's first version aids the application scientists in writing code for accelerators like GPUs.
This Adviser, however, can be extended to support all OpenMP-enabled devices.

The tool has two modes: Training mode and Prediction mode.
The training mode must be executed on the target hardware.
It takes all benchmark codes as input and generates all possible variants of the code to run on the target device.
Then it collects data from all generated codes to train an ML-based cost model for use in prediction mode.
In prediction mode the tool does not need any interaction with the target device.
It accepts C/C++ code as input and returns the best code variant that can be used to offload the code to the specified device.
The tool can determine the kernels that are best suited for offloading by predicting their runtime using a machine learning-based cost model as defined in Section~\ref{subsec:costmodel}.

\subsection{Attribubtes}
The following are the key attributes of  the OpenMP Advisor.
\begin{enumerate}[nosep,leftmargin=*]
    \item \textbf{Static} -- Access to GPUs is a significant challenge when analyzing an application using GPU offloading. 
    During development, many application scientists may not have easy access to GPUs. 
    As a result, it is critical that in the prediction mode our Advisor performs all its analysis at compile time and does not require runtime profiling. 
    Even in the training mode, sometimes it is impractical to predict certain values during code generation.
    In order to help the advisor in such circumstances, we provide a json file for input where users can provide these values.
    
    \item \textbf{Minimalistic} -- The Advisor makes no changes to the kernel's body. 
    It simply identifies the application's \texttt{omp target} regions and adds various OpenMP directives and clauses to generate multiple kernel variants.
    
    \item \textbf{Correctness} -- 
    The Advisor verifies the correctness of the generated code to keep in line with OpenMP programming model.
    Currently, the advisor does not modify the kernel body and does not guarantee the correctness of its content. 
    Consequently, despite its ability to predict the optimal scenario for GPU offloading, it generates the top N (N provided by the user) code variants and lets the application scientist select which code to utilize.
    
    \item \textbf{Portability} -- The Advisor is portable to multiple compiler and HPC clusters. 
    In our work we worked on four different compilers: LLVM/clang (nvptx), LLVM/clang (rocm), GNU/gcc and NVIDIA HPC/nvc. 
    The advisor also works for a variety of HPC clusters with different GPUs: such as Summit~\cite{Summit}, Ookami~\cite{burford2021ookami}, Wombat~\cite{Wombat} and Seawulf~\cite{Seawulf} clusters using NVIDIA GPUs and Corona~\cite{Corona} using AMD GPUs.
    Limited work was also done on the Intel Devclouds~\cite{Intel-Devcloud} with Intel GPU and icpx compiler.
    
    \item \textbf{Adaptability} -- The advisor is adaptable enough to accept new applications.
    When attempting to predict for a new application where real-time data collection is difficult or impractical, the application scientist could create a proxy application similar to the target application and train the model using data collected on the proxy application.
\end{enumerate}

\subsection{Design}
Before proceeding with the implementation, there were a few design challenges that needed to be addressed.
The first concern was which toolkit or compiler to employ in the development of our framework.
This was a straightforward question since we used the most popular library at the time -- the LLVM compiler project.
LLVM is a library for building, optimizing, and producing intermediate and/or binary machine code.
Like most modern compilers, LLVM divides the compilation process into three levels:
\begin{itemize} [nosep,leftmargin=*]
    \item \textbf{Front-end} translates high-level language (like C/C++, fortran) to LLVM-IR.
    \item \textbf{Middle-end} performs several optimization passes on the LLVM-IR.
    \item \textbf{Back-end} converts the optimized LLVM-IR into the assembly language of the corresponding platform.
\end{itemize}

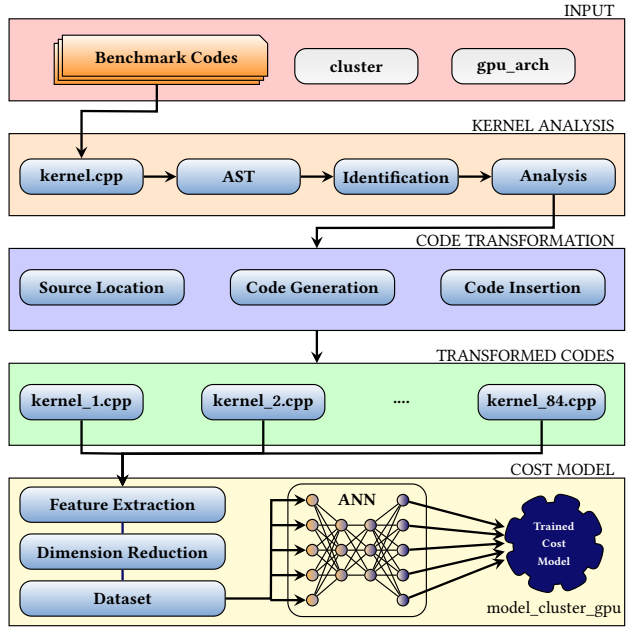
\begin{figure}[b!]
    \vspace{-2mm}
    \resizebox {\columnwidth} {!} {
\centering
\begin{tikzpicture}

    \draw[color=black, fill = red!20] (0.25, 7.7) rectangle(7.75, 6.7);
    \node[empty, anchor=east] at (7.75, 7.8) {INPUT};
    
	\node[empty](doc) at (0.8, 6.9) {};
    \foreach \i in {0, 1, 2} { 
        \draw[bottom color=orange!80, top color=orange!5]  ($(doc)+(.05*\i, .05*\i)$) -- ++(2.5, 0) -- ++(0, 0.4)  -- ++(-0.1, 0.1) -- ++(-2.4, 0) -- cycle;
    };
    \node[empty, minimum width=3.5, above right=0.21 and 0.25 of doc, anchor=west] { \textbf{Benchmark Codes} };
    
    \node[style4, right=2.8 of doc, anchor=south west] (cluster) {\textbf{cluster}};
    \node[style4, right=0.4 of cluster.east, anchor=west] (gpu) {\textbf{gpu\_arch}};

    \draw[color=black, fill = orange!20] (0.25, 6.3) rectangle(7.75, 5.3);
    \node[empty, anchor=east] at (7.75, 6.4) {KERNEL ANALYSIS};
    
    \node[style1, below left=1cm and 0.3cm of doc, anchor=west] (kernel) {\textbf{kernel.cpp}};
    \node[style1, right=0.4 of kernel](AST) {\textbf{AST}};
    \node[style1, right=0.4 of AST](identify) {\textbf{Identification}};
    \node[style1, right=0.4 of identify, anchor=west](analysis) {\textbf{Analysis}};

    \draw[myarrows] ($(doc)+(1.25cm,0)$) -- ($(doc)+(1.25cm,-0.33cm)$) -- ($(kernel)+(0,0.8cm)$) -- (kernel);
    \draw[myarrows] (kernel) -- (AST);
    \draw[myarrows] (AST) -- (identify);
    \draw[myarrows] (identify) -- (analysis);
    
    \draw[color=black, fill = blue!20] (0.25, 4.9) rectangle(7.75, 3.9);
    \node[empty, anchor=east] at (7.75, 5.0) {CODE TRANSFORMATION};
    
    \node[style1, minimum width=2cm, below=1.35cm of kernel.west, anchor=west] (loc) {\textbf{Source Location}};
    \node[style1, minimum width=2cm, right=0.55 of loc](code) {\textbf{Code Generation}};
    \node[style1, minimum width=2cm, right=0.55 of code](insert) {\textbf{Code Insertion}};

    \draw[myarrows] (analysis) -- ($(analysis.south)+(0,-0.4)$) -- (4, 5.15) -- (4, 4.9);
    
    \draw[color=black, fill = green!20] (0.25, 3.5) rectangle(7.75, 2.5);
    \node[empty, anchor=east] at (7.75, 3.6) {TRANSFORMED CODES};
    \node[style1, below=1.4cm of loc.west, anchor=west] (var1) {\textbf{kernel\_1.cpp}};
    \node[style1, right=0.7cm of var1, anchor=west] (var2) {\textbf{kernel\_2.cpp}};
    \node[empty, right=0.7cm of var2, anchor=west] (var3) {\textbf{....}};
    \node[style1, right=0.7cm of var3, anchor=west] (var84) {\textbf{kernel\_84.cpp}};
        
    \draw[myarrows] (4, 3.9) -- (4, 3.5);
    
    \draw[color=black, fill = yellow!20] (0.25, 2.1) rectangle(7.75, 0.3);
    \node[empty, anchor=east] at (7.75, 2.2) {COST MODEL};
    
    \node[style1, minimum width=2.5cm, below=1.25cm of var1.west, anchor=west] (feature) {\textbf{Feature Extraction}};
    \node[style1, minimum width=2.5cm, below=0.35cm of feature.south west, anchor=west] (dimension) {\textbf{Dimension Reduction}};
    \node[style1, minimum width=2.5cm, below=0.35cm of dimension.south west, anchor=west] (dataset) {\textbf{Dataset}};
    
    \draw[myarrows] (var1) -- ($(var1.south)-(0,0.4)$) -- ($(feature.north)+(0,0.4)$) -- (feature);
    \draw[myarrows] (var2) -- ($(var2.south)-(0,0.4)$) -- ($(feature.north)+(0,0.4)$) -- (feature);
    \draw[myarrows] (var84) -- ($(var84.south)-(0,0.4)$) -- ($(feature.north)+(0,0.4)$) -- (feature);

    \draw[thick, dkblue] (feature) -- (dimension);
    \draw[thick, dkblue] (dimension) -- (dataset);
    
	\begin{scope}[shift={($(dimension.east)+(4,0.1)$)}]
        \draw[base, fill=dkblue] \gear{8}{0.5}{0.6}{10}{2}{TrainedModel}{\textcolor{white}{\textbf{Trained \\Cost Model}}};
    \end{scope}
    \node[empty, below=0.2 of TrainedModel] {model\_cluster\_gpu};
    
    \node[smallcir2, above right=0.35 and 1 of dimension] (in1) {};
    \node[smallcir2, below=0.15 of in1] (in2) {};
    \node[smallcir2, below=0.15 of in2] (in3) {};
    \node[smallcir2, below=0.15 of in3] (in4) {};
    \node[smallcir2, below=0.15 of in4] (in5) {};
    
    \node[smallcir3, right=0.2 of in2] (h11) {};
    \node[smallcir3, right=0.2 of in3] (h12) {};
    \node[smallcir3, right=0.2 of in4] (h13) {};
    
    \node[smallcir4, right=0.2 of h11] (h21) {};
    \node[smallcir4, right=0.2 of h12] (h22) {};
    \node[smallcir4, right=0.2 of h13] (h23) {};
    
    \node[smallcir5, right=0.95 of in1] (out1) {};
    \node[smallcir5, right=0.95 of in2] (out2) {};
    \node[smallcir5, right=0.95 of in3] (out3) {};
    \node[smallcir5, right=0.95 of in4] (out4) {};
    \node[smallcir5, right=0.95 of in5] (out5) {};

    \node[base, rounded corners=5, above left=0.15 and 0.25 of in1, minimum width = 48, minimum height = 46, anchor=north west](nn) {};
    \node[empty, below=0 of nn.north] {\textbf{ANN}};
    
    \draw[myarrows] (dataset.east) -- ($(dataset.east)+(0.55,0)$) -- ($(in1)-(0.5,0)$) -- (in1);
    \draw[myarrows] ($(in2)-(0.5,0)$) -- (in2);
    \draw[myarrows] ($(in3)-(0.5,0)$) -- (in3);
    \draw[myarrows] ($(in4)-(0.5,0)$) -- (in4);
    \draw[myarrows] ($(dataset.east)+(0.55,0)$) -- ($(in5)-(0.5,0)$) -- (in5);
    
    \foreach \i in {1,2,3,4,5} {
        \draw[thin] (in\i) -- (h11);
        \draw[thin] (in\i) -- (h12);
        \draw[thin] (in\i) -- (h13);
    }
    \foreach \i in {1,2,3} {
        \draw[thin] (h1\i) -- (h21);
        \draw[thin] (h1\i) -- (h22);
        \draw[thin] (h1\i) -- (h23);
    }
    \foreach \i in {1,2,3,4,5} {
        \draw[thin] (h21) -- (out\i);
        \draw[thin] (h22) -- (out\i);
        \draw[thin] (h23) -- (out\i);
    }
    
    \foreach \i\j in {1/-0.2,2/-0.1,3/0,4/0.1,5/0.2} {
        \draw[myarrows] (out\i) -- ($(TrainedModel)-(0.6,\j)$);
    }
    
\end{tikzpicture}
}
	\caption{High level flow of the Training mode}
	\label{fig:flow_train}
\end{figure}

We chose the Clang compiler (ver 14.0.0) because our target applications are written in C/C++.
Clang is a tool development platform as well as the front-end for the C language family in the LLVM project.
Because of its library-based architecture, it is easier and more flexible to reuse and integrate new features into other projects.

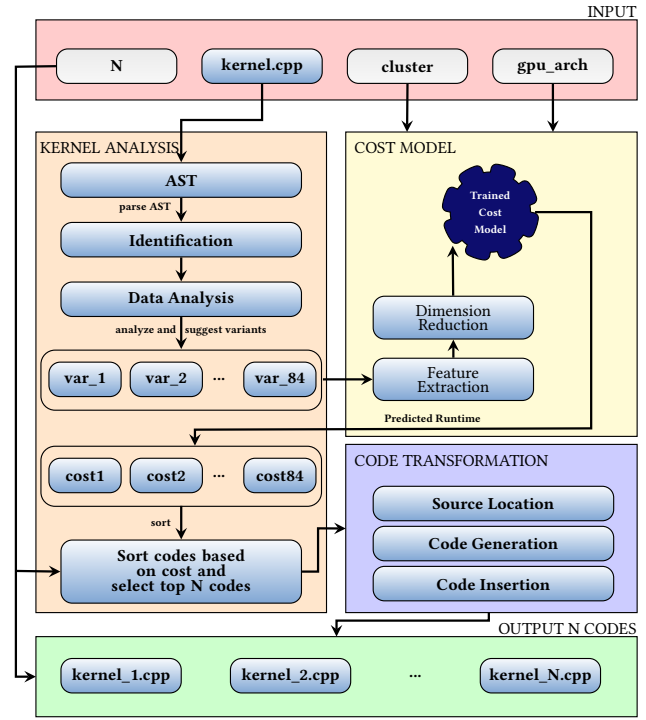
\begin{figure}[b!]
\vspace{-2mm}
    \resizebox {\columnwidth} {!} {
\centering
\begin{tikzpicture}

    \draw[color=black, fill = red!20] (0.25, 9.0) rectangle(7.75, 8.0);
    \node[empty, anchor=east, inner sep = 0] at (7.75, 9.1) {INPUT};
    
	\node[style4, anchor=south west](num) at (0.5, 8.2) {\textbf{N}};
    \node[style1, right=0.3 of num.east, anchor=west] (kernel) {\textbf{kernel.cpp}};
    \node[style4, right=0.3 of kernel, anchor=west] (cluster) {\textbf{cluster}};
    \node[style4, right=0.3 of cluster.east, anchor=west] (gpu) {\textbf{gpu\_arch}};
    
    \draw[color=black, fill = orange!20] (0.25, 7.6) rectangle(3.875, 1.6);
    \node[empty, anchor=west] at (0.18, 7.4) {KERNEL ANALYSIS};
    
    \node[style1, minimum width= 3cm](AST) at (2.0625, 7.0) {\textbf{AST}};
    \node[style1, minimum width= 3cm, below=0.3 of AST](identify) {\textbf{Identification}};
    \node[style1, minimum width= 3cm, below=0.3 of identify](analysis) {\textbf{Data Analysis}};

    \draw[myarrows] (kernel) -- ($(kernel.south)-(0,0.45)$) -- ($(AST)+(0,0.75)$) -- (AST);
    \draw[myarrows] (AST) -- (identify) node[arrow, pos=0.4, left] {\textbf{parse AST}};
    \draw[myarrows] (identify) -- (analysis);
    
    \node[style1, minimum width=0.9cm, below left=1 and 0.15 of analysis.west, anchor=west](var1) {\textbf{var\_1}};
    \node[style1, minimum width= 0.9cm, right=0.1 of var1, anchor=west](var2) {\textbf{var\_2}};
    \node[empty, right=0 of var2, anchor=west, align=center](var3) {\textbf{...}};
    \node[style1, minimum width= 0.9cm, right=0.1 of var3, anchor=west](var84) {\textbf{var\_84}};
    
    \draw[myarrows] (analysis) -- ($(analysis.south) - (0, 0.4)$) node[arrow, pos=0.4, right=-0.95] {\textbf{analyze and { } suggest variants}};

    \node[style1, minimum width= 0.9cm, below=1.2 of var1.west, anchor=west](cost1) {\textbf{cost1}};
    \node[style1, minimum width= 0.9cm, right=0.1 of cost1, anchor=west](cost2) {\textbf{cost2}};
    \node[empty, right=0 of cost2, anchor=west, align=center](cost3) {\textbf{...}};
    \node[style1, minimum width= 0.9cm, right=0.1 of cost3, anchor=west](cost84) {\textbf{cost84}};
    
    \node[base, rounded corners=5, above left=0.15 and 0.1 of var1, minimum width = 3.5cm, minimum height = 0.75cm, anchor=north west](variants) {};
    \node[base, rounded corners=5, above left=0.15 and 0.1 of cost1, minimum width = 3.5cm, minimum height = 0.75cm, anchor=north west](variants) {};
    
    \node[style1, below=3.4 of analysis.west, anchor=west, minimum width=3cm, text width=2cm, align=center](sort) {\textbf{Sort codes based on cost and \\select top N codes}};
    
    \draw[myarrows] ($(sort.north)+(0,0.42)$) -- (sort) node[arrow, pos=0.5, left] {\textbf{sort}};
    
    \draw[color=black, fill = yellow!20] (4.115, 7.6) rectangle(7.75, 3.8);
    \node[empty, anchor=west](modelText) at (4.1, 7.4) {COST MODEL};
    
	\begin{scope}[shift={($(AST.east)+(2.35,-0.4)$)}]
        \draw[base, fill=dkblue] \gear{8}{0.5}{0.6}{10}{2}{TrainedModel}{\textcolor{white}{\textbf{Trained \\Cost Model}}};
    \end{scope}
    
    \draw[myarrows] (cluster.south) -- (4.88, 7.6);
    \draw[myarrows] (gpu.south) -- (6.7, 7.6);
    
    \node[style1, right=0.7 of var84, anchor=west, minimum width=2cm, text width=1cm, align=center](feature) {Feature Extraction};
    \node[style1, above=0.5 of feature.north west, minimum width=2cm, anchor=west, text width=1cm, align=center](dimension) {Dimension Reduction};

    \draw[myarrows] ($(var84.east)+(0.08,0)$) -- (feature);
    \draw[myarrows] (feature) -- (dimension);
    \draw[myarrows] (dimension) -- ($(TrainedModel)-(.47,0.42)$);
    \draw[myarrows] ($(TrainedModel.east)-(0.2,0)$) -- ($(TrainedModel.east)+(0.5,0)$) -- ($(TrainedModel.east)+(0.5,-2.7)$) -- ($(cost2.north east)+(-0.1,0.35)$) node[arrow, pos=0.4, above=-0.05] {\textbf{Predicted Runtime}} -- ($(cost2.north east)+(-0.1,0.15)$);

    \draw[color=black, fill = blue!20] (4.115, 3.7) rectangle(7.75, 1.6);
    
    \draw[myarrows] (sort.east) -- ($(sort.east) + (0.2, 0)$) -- ($(sort.east) +(0.2, 0.54)$) -- (4.115, 2.65);
    
    \node[empty, anchor=west] at (4.1, 3.5) {CODE TRANSFORMATION};
    \node[style1, below=1.55 of feature.west, anchor=west, minimum width=3cm, align=center](loc) {\textbf{Source Location}};
    \node[style1, below=0.5 of loc.west, anchor=west, minimum width=3cm, align=center](gen) {\textbf{Code Generation}};
    \node[style1, below=0.5 of gen.west, anchor=west, minimum width=3cm, align=center](insert) {\textbf{Code Insertion}};

    \draw[color=black, fill = green!20] (0.25, 1.3) rectangle(7.75, 0.3);
    \node[empty, anchor=east, inner sep = 0] at (7.75, 1.4) {OUTPUT N CODES};
    
    \draw[myarrows] (0, 2.12) -- (sort.west);
    \draw[myarrows] (num.west) -- (0, 8.415) -- (0, 0.8) -- (0.25, 0.8);
    \draw[myarrows] (5.9, 1.6) -- (5.9,1.5) -- (4.0, 1.5) -- (4.0, 1.3);
    
    \node[style1, text width=1.5cm, minimum width=1.5cm, inner sep=0pt, align=center, below=1.3 of sort.west, anchor=west] (kern1) {\textbf{kernel\_1.cpp}};
    \node[style1, right=0.6 of kern1, text width=1.5cm, minimum width=1.5cm, inner sep=0pt, align=center,anchor=west] (kern2) {\textbf{kernel\_2.cpp}};
    \node[empty, right=0.6 of kern2, align=center,anchor=west] (kern3) {\textbf{...}};
    \node[style1, right=0.6 of kern3, text width=1.5cm, minimum width=1.5cm, inner sep=0pt, align=center,anchor=west] (kernN) {\textbf{kernel\_N.cpp}};
\end{tikzpicture}
}
	\caption{High level flow of the Prediction mode}
	\label{fig:flow}
\end{figure}

We then had to decide at what level we should implement our framework.
Despite the fact that LLVM's strength lies in the LLVM-IRs, our requirement is to generate and return C/C++ code to the scientist.
We require the accurate source location from a C/C++ file in order to insert OpenMP directives into that file.
Although the LLVM-IR can contain source code information in source level debugging, the accuracy of the source location cannot be guaranteed.
On the other hand Clang's AST closely resembles both the written C++ code and the C++ standard.
Clang has a one-to-one mapping of each token to the AST node and an excellent source location retention for all AST nodes.
The AST is the best place to get accurate source code information.
Hence we implemented our Advisor in Clang.
The Advisor follows the design depicted in Figure~\ref{fig:flow_train} in the training mode, and  Figure~\ref{fig:flow} in the prediction mode.
Both modes have three major modules, which are explained in the following subsections - \textit{Kernel Analysis} (Section~\ref{subsec:analysis}), \textit{Cost Model} (Section~\ref{subsec:costmodel}) and \textit{Code Transformation} (Section~\ref{subsec:transform}).

\section{Implementation}
\label{sec:impl}

A preprocessed AST is much easier to work with than source-level C/C++ code, and we can always easily refer to the original code by using Clang's Plugins~\cite{clangplugin} interface.
Clang Plugins enable the execution of additional user-defined actions during compilation.
The compiler loads a plugin from a dynamic library at runtime.
The \texttt{FrontendAction} interface, which enables user-specific actions to be executed as part of the compilation, is the typical entry point to a Clang Plugin.
To run the tools over the AST, Clang provides the \texttt{ASTFrontendAction} interface, that handles action execution.
We define and register \texttt{PluginOMPAdvisor}, a plugin that implements the \texttt{CreateASTConsumer} method, which returns an \texttt{OMPAdvisorASTConsumer} that consumes (or reads) the Clang AST.
We can create generic actions on an AST using this \texttt{ASTConsumer} interface.
This interface provides the \texttt{HandleTranslationUnit} function, which is only called after the entire source file has been parsed.
In this case, a translation unit effectively represents an entire source file, and an \texttt{ASTContext} object represents the AST for that source file.

Clang also provides a \texttt{RecursiveASTVisitor} class, which traverses the entire Clang AST in preorder or postorder depth-first order, visiting each node.
In our Advisor we define an \texttt{OMPAdvisorVisitor}, which is an instance of the \texttt{RecursiveASTVisitor}, to visit any type of AST node, such as \texttt{FunctionDecl} and \texttt{Stmt}, by simply overriding their visit function, such as \texttt{VisitStmt} or \texttt{VisitFunctionDecl}.
Rather than directly calling any of the \texttt{Visit*} functions, we use the \texttt{TraverseDecl} function, which calls the appropriate \texttt{Visit*} function in the background.
To continue traversing the AST (to investigate additional nodes), we return \texttt{true} from such \texttt{Visit*} functions, and \texttt{false} to stop the traversal and effectively terminate the Clang compilation process.

\subsection{Kernel Analysis}
\label{subsec:analysis}
This is the first module which the \texttt{HandleTranslationUnit} method calls in both the Prediction and Training mode.
As the name implies, this module analyzes an OpenMP kernel.
Overall, this module takes as input a C/C++ source file, analyzes it, and outputs all possible GPU offloading variants.
This module is responsible for three tasks: \textit{Identifying Kernels, Data Analysis} and \textit{Variant Generation}.

\subsubsection{Identifying Kernels.}
User's input is utilized to help us identify a target region since the scope of this project is not to automatically parallelize the code.
An application scientist only needs to use the ``\texttt{omp target} directive'' to mark the region.
We parse the clang AST to search for the \texttt{OMPTargetDirective} node, which is a subclass of the \texttt{OMPExecutableDirective} class and an instance of the \texttt{Stmt} class.
We override the \texttt{OMPAdvisorVisitor} class's \texttt{VisitStmt} method and check each visited \texttt{Stmt} to see if they are the \texttt{OMPTargetDirective} node.
Once a kernel is identified, we assign it a unique id and create an instance of the class "KernelInformation."
In that object we store information like, unique\_id, start and end locations of the kernel, function from which the kernel is called, whether the kernel is called from within a loop and the number of nested \texttt{for} loops.
All of this information will further be used to identify variables accessed within kernels.

\subsubsection{Data Analysis}
The next step is to determine what data the kernel uses.
Due to the high cost of data transfers both to and from the GPU, careful data management and mapping between host and device are required for the use of accelerators in HPC.
OpenMP does not specify how the data should be handled by the compiler in implicit data transfer.
Application scientists are deterred from GPU utilization due to the complexities in achieving effective data management.
Automatically identifying and utilizing GPU data in an application can reduce the burden on application scientists and improve overall execution time.
We expanded on Mishra et al's work~\cite{mishra2020data} on \textit{Data Transfer and Reuse Analysis}, which describes how to determine which variables are being used by a kernel.
We use the Clang AST to implement \textit{live variable analysis} for each kernel, concentrating only on variables used within a kernel.
One limitation of this work is that it cannot determine the range of an array during compilation.
In our implementation, we extended the analysis to determine the array's range as declared in the code.
However, such an analysis is not always possible, particularly when dealing with global variables or pointer manipulation.
To overcome this obstacle we expect the user to specify the array's range in the \texttt{data map} clause.
If the user does not specify the range of an array to map to the GPU, we assume that all arrays used within the kernel are declared in or passed to the kernel-defining function.
We continue to perform live variable analysis to determine which variables to synchronize between the CPU and GPU and calculate their sizes.
Our approach has a limitation in that we currently map data between the CPU and GPU only before and after the kernel.
Managing data transfer during kernel execution is a task for the future.

Before the variables are stored in the \texttt{KernelInformation} object, we classify them into five groups, based on how they are accessed before, during, or after the kernel:
\begin{itemize}[nosep,leftmargin=*]
\item \textbf{\textit{alloc}}: These are the variables that are being \textit{assigned within} the kernel for the first time and no data transfer to the GPU is required.

\item \textbf{\textit{to}}: These are variables that were \textit{assigned before} but were only \textit{accessed} and not modified within the kernel.
This data must be transferred to the device but does not have to be transferred back to the host.

\item \textbf{\textit{from}}: These are variables that are \textit{updated within} the kernel and \textit{accessed after} the kernel call.
This category does not require data transfer to the device, but it does necessitate data transfer from the device to the host.

\item \textbf{\textit{tofrom}}: These are variables that are \textit{assigned before}, \textit{updated within} and \textit{accessed after} the kernel definition.
This type of data must be transferred in both directions between the host and the device.

\item \textbf{\textit{private}}: Finally, these are variables that are defined and used \textit{only within} the kernel, and does not need to be transferred between the host and the device.
\end{itemize}

Data labeled \textit{alloc}, \textit{to}, and \textit{tofrom} are mapped in ``\texttt{omp target enter data}'' directives before the kernel, while data labeled \textit{from} and \textit{tofrom} are mapped in ``\texttt{omp target exit data}'' directives after the kernel.

\subsubsection{Variant Generation}
\label{subsec:varGen}
\begin{figure}[t!]
\begin{lstlisting}[caption={Four nested ``\texttt{for}'' loops, with ``\texttt{teams distribute}'' in Loop $0$ and ``\texttt{parallel for collapse(3)}'' in Loop $1$.}, label=lst:advisor:for_loops,frame=tlrb, captionpos=b]
<@\color{mauve}{\#pragma omp target teams distribute collapse(1)}@>
  for (int i = 0; i < N_i; i++) {    <@\color{dkgreen}{<= Loop 0}@>
<@\color{mauve}{\#pragma omp parallel for collapse(3) schedule(static) }@>
    for (int j = 0; j < N_j; j++) {    <@\color{dkgreen}{<= Loop 1}@>
      for (int k = 0; k < N_k; k++) {    <@\color{dkgreen}{<= Loop 2}@>
        for (int l = 0; l < N_l; l++) {    <@\color{dkgreen}{<= Loop 3}@>
          ...
        }
      }
    }
  }
\end{lstlisting}
\vspace{-8mm}
\end{figure}
Finally, we generate a number of different kernel variants that can be used to offload the kernel to the GPU.
We'll start by counting how many nested collapsible for loops are there.
In the current implementation, we can check up to four levels of collapsing the for loops.
We chose four nested loops because, as shown in Code~\ref{lst:advisor:dslash}, the Wilson Dslash kernel has four for loops.
Each of these for loops is given a unique Loop number ranging from $0 - 3$.
Loop 0 is always expected to be distributed across all teams on the GPU.
The variants are generated based on five parameter:
\begin{itemize}[leftmargin=*]
    \item Value of collapse used in \texttt{distribute} directive
    \item Value of collapse used in \texttt{parallel for} directive
    \item Position of the \texttt{parallel for} directive
    \item Scheduling type of the loop iterations
    \item Data transfer between the host and the device
\end{itemize}

\begin{table}[b!]
\vspace{-4mm}
\centering
\renewcommand{\arraystretch}{1.5}
\begin{tabular}{|c|c|}
  \hline
  \rowcolor{black}
  \textcolor{white}{\textbf{\# for loops}} & \textcolor{white}{\textbf{\# Variants}}\\[0.25em]
  \hline
  1 & 6 \\ \hline
  2 & 18 \\ \hline
  3 & 42 \\ \hline
  4 & 84 \\ \hline
\end{tabular}
\caption{Number of Variants expected to be generated for each number of for loops}
\label{tab:advisor:variants}
\end{table}
\begin{table*}[t]
\small
\begin{tabular}{|L{0.16\textwidth}|C{0.12\textwidth}|L{0.65\textwidth}|}
\hline
\rowcolor{black}
\textcolor{white}{\textbf{Application}} & \textcolor{white}{\textbf{Domain}} & \textcolor{white}{\textbf{Description}} \\ \hline
\textbf{Breadth First Search} (bfs)~\cite{bfs-rodinia}$^{\dagger}$ & Graph Algorithms & BFS is a graph traversal algorithm frequently employed in a variety of academic and professional fields.
\\ \hline

\textbf{Correlation Coefficient} (correlation)~\cite{schober2018correlation} & Statistics & The Pearson's Correlation Coefficient is a measurement of the linear correlation between two sets of data. \\ \hline

\textbf{Covariance} (covariance) & Probability Theory & The sample covariance are statistics calculated from a sample of data on one or more random variables.  \\ \hline

\textbf{Gaussian Elimination} (gauss)~\cite{gauss-rodinia}$^{\dagger}$ & Linear Algebra & Gaussian Elimination computes result row by row, solving for all of the variables in a linear system.
\\ \hline

\textbf{K-nearest neighbors} (knn)~\cite{knn-rodinia}$^{\dagger}$ & Data Mining & The k-nearest neighbors technique utilizes proximity to classify or predict how a single data point will be grouped. 
It is a non-parametric, supervised learning classifier. \\ \hline

\textbf{Laplace's Equation} (laplace)~\cite{kedia2009hybrid} & Numerical Analysis & In physics, Laplace's equation is a second-order partial differential equation. \\ \hline

\textbf{Matrix-Matrix Multiplication} (mm) & Linear Algebra & The most significant matrix operation is arguably matrix multiplication, extensively utilized in a variety of fields, including the solution of linear systems of equations, translation of coordinate systems, etc. \\ \hline

\textbf{Matrix Vector Multiplication} (mv) & Linear Algebra & A simple matrix-vector multiplication. \\ \hline

\textbf{Matrix Transpose} (mt) & Linear Algebra & The transpose of a matrix is simply a flipped version of the original matrix. \\ \hline

\textbf{Particle Filter} (particle)~\cite{particlefilter-rodinia}$^{\dagger}$ & Medical Imaging & The particle filter is a statistical estimator of a target item's location given noisy measurements of the target's location and an idea of the object's movement in a Bayesian framework. \\ \hline

\textbf{Proxy App} (proxy\_app) & -NA- & This is a proxy app that has same number of loops and performs similar computation to our target app, the Wilson Dslash operator. 
Whenever it is difficult to collect data on real applications, proxy apps help us collect more data.
\\ \hline

\end{tabular}
\caption{Benchmark Applications. Apps marked $^{\dagger}$  are adopted from the Rodinia Benchmark Suite~\cite{che2009rodinia}.}
\label{ompadv:tab:bench}
\vspace{-4mm}
\end{table*}

The total number of \texttt{for} loops and the position of the \texttt{parallel for} directive determine the maximum value of \texttt{collapse} that can be used in the \texttt{teams distribute} and \texttt{parallel for} directive.
Suppose there are four \texttt{for} loops, as shown in Code~\ref{lst:advisor:for_loops}.
If the \texttt{parallel for} directive is at Loop 0, the ``\texttt{teams distribute parallel for}" directive will be combined and thus the \texttt{collapse} clause for \texttt{teams distribute} directive does not exist.
If the \texttt{parallel for} directive is located on Loop $x$ (where $1 \leq x \leq 3$ ), then the maximum possible value of collapse for the \texttt{teams distribute} directive is $x$.
Whereas for the \texttt{parallel for} directive, the maximum possible value of \texttt{collapse} is $(NUM-x)$, where $NUM$ is the total number of for loops.

The scheduling type of the loop iteration could be one of \texttt{static}, \texttt{dynamic} or \texttt{guided}.
Using different permutations of these parameters, we could generate a variety of GPU offloading code variants. 
Once all of the variants have been generated, we use our static cost model to predict the runtime of each of these generated kernels.
The total number of variants that can be produced for a particular number of for loops is shown in Table~\ref{tab:advisor:variants}.

\subsection{Cost Model}
\label{subsec:costmodel}
A compile-time cost model is required to select the best option from all the variants generated by the Kernel Analysis module.
Most modern compilers employ analytical models to calculate the cost of executing the original code as well as the various offloading code variants.
Building such an analytical model for compilers is a difficult task that necessitates a lot of effort on the part of a compiler engineer. 
Recently, machine learning techniques have been successfully applied to build cost models for a variety of compiler optimization problems.
For our tool we extended COMPOFF~\cite{mishra2022compoff} to be used as our cost model.
COMPOFF is a machine learning based compiler cost model that statically estimates the cost of OpenMP offloading using an artificial neural network model.
Their results show that this model can predict offloading costs with an average accuracy greater than 95\%.
The major limitation of COMPOFF was that they had to train a separate model for each variant.
In our work, we add more training data and extend it to train a single cost model for all variants.

As soon as we know the prediction for the generated variant, we store it in the instance of the \texttt{KernelInformation} class so that the Kernel Transformation module can use it.
But the biggest challenge in implementing an ML based cost model is the lack of available training data.
To overcome this problem, we wrote additional benchmark applications and adopted some benchmarks from the Rodinia benchmark suite~\cite{che2009rodinia}.
The goal is to include a broader class of benchmarks that would cover the spectrum of statistical simulation, linear algebra, data mining, etc.
For this we developed applications tabulated in Table~\ref{ompadv:tab:bench}.
More applications from various other domains will be added to this repository in the future.

\subsection{Kernel Transformation}
\label{subsec:transform}
In the Kernel Transformation module we need to actually transform the original source code based on the analysis and predictions from the previous modules.
For the given kernel, we generate every possible code variation in the Training mode.
However, before we can generate code in Prediction mode, we must first address another crucial question.
Which code should we generate?
Should we only generated code for the fastest kernel? 
Regrettably, once the directives are in place, neither the Advisor nor OpenMP validate the kernel's correctness. 
This is inline with the OpenMP philosophy as well.
As a result, we can only guarantee the correctness of the generated OpenMP directive in our framework.

So how can we overcome this problem?
We could generate code for every possible variation, as we do during training, and let the user choose which one to use.
But this means that users will be overwhelmed with information.
Alternatively, we could ask the user to provide a number for the maximum number of codes to generate.
The predicted runtime can be put as a comment before the kernel in every piece of code.
The application scientist will then have more power to accept or reject the generated code.
We will be able to produce a single code and provide it to the user once the issue of validating an OpenMP code for correctness is resolved.
Until then, our Advisor will be able to generate the top best variants as specified by the application scientist.

Regardless, we need to write a module to modify the existing source code and generate a new code.
Clang provides the \texttt{Rewriter} interface, whose primary function is to route high-level requests to the involved low-level \texttt{RewriteBuffers}.
A \texttt{Rewriter} assists us in managing the code rewriting task.
In the \texttt{Rewriter} interface we can set the \texttt{SourceManager} object which handles loading and caching of source files into memory. 
This object assigns distinct FileIDs to each distinct \texttt{\#include} chain and is the owner of the \texttt{MemoryBuffer} objects for all of the loaded files.
The \texttt{SourceManager} can be queried to obtain information about \texttt{SourceLocation} objects, which can then be converted into spelling or expansion locations.
Spelling locations represent the bytes that correspond to a token, while expansion locations represent the location in the source file of the code.
For example, in the case of a macro expansion, the spelling location indicates where the expanded token originated, while the expansion location specifies where it was expanded.
\begin{figure}[b!]
\begin{lstlisting}[caption={Location of the seven generated code.}, label=lst:advisor:code_gen,frame=tlrb, captionpos=b, aboveskip=-0.5\baselineskip]
0: <@\color{dkgreen}{// Predicted Runtime: 1.2 s }@>
1: <@\color{red}{\#pragma omp target enter data map(...)}@>
2: <@\color{mauve}{\#pragma omp target teams distribute  collapse(2)}@>
    for (int i = 0; i < N_i; i++) {
3:    
      for (int j = 0; j < N_j; j++) {
4: <@\color{mauve}{\#pragma omp parallel for schedule(dynamic) }@>
        for (int k = 0; k < N_k; k++) {
5: 
          for (int l = 0; l < N_l; l++) {
            ...
          }
        }
      }
    }
6: <@\color{red}{\#pragma omp target exit data map(...)}@>
\end{lstlisting}
\end{figure}

The \texttt{Rewriter} is a critical component of our plugin's kernel transformation module.
The strategy used here is to meticulously alter the original code at crucial locations to carry out the transformation rather than handling every possible AST node to spit back code from the AST.
For this, the \texttt{Rewriter} interface is essential. 
It is an advanced buffer manager that effectively slices and dices the source using a rope data structure.
For each \texttt{SourceManager}, the \texttt{Rewriter} also stores the low-level \texttt{RewriteBuffer}.
Together with Clang's excellent retention of source location for all AST nodes, \texttt{Rewriter} makes it possible to remove and insert code very precisely in the \texttt{RewriteBuffer}. 
When the update is finished, we can dump the \texttt{RewriteBuffer} into a file to obtain the updated source code.

\begin{table}[b!]
\centering
\small
\begin{tabular}{|C{0.145\columnwidth}|C{0.19\columnwidth}|C{0.35\columnwidth}|C{0.135\columnwidth}|}
  \hline \rowcolor{black}
  \textcolor{white}{\small \textbf{Cluster}} & \textcolor{white}{\small \textbf{GPU}} &
  \textcolor{white}{\small \textbf{Compiler}} &
  \textcolor{white}{\small \textbf{Version}} \\
  \hline
  \textbf{Summit} &  & LLVM/clang (nvptx) & 13.0.0 \\ \cline{3-4}
  \cite{Summit} & \multirow{-2}{0.15\columnwidth}{\centering \scriptsize{\textbf{NVIDIA Tesla V100}}} & GNU/gcc (nvptx-none) & 9.1.0 \\ \hline
  \textbf{Corona} \cite{Corona} & \scriptsize{\textbf{AMD Radeon Instinct MI50}} & LLVM/clang (rocm-5.3) & 15.0.0 \\ \hline
  \textbf{Ookami} \cite{burford2021ookami} & \scriptsize{\textbf{NVIDIA Tesla V100}} & LLVM/clang (nvptx) & 14.0.0 \\ \hline
  \textbf{Wombat} \cite{Wombat} & \scriptsize{\textbf{NVIDIA Tesla A100}} & LLVM/clang (nvptx) & 15.0.0 \\ \hline
  \textbf{Seawulf} &  & LLVM/clang (nvptx) & 12.0.0 \\ \cline{3-4}
  \cite{Seawulf} & \multirow{-2}{0.15\columnwidth}{\centering \scriptsize{\textbf{NVIDIA Tesla K80}}} & NVIDIA/NVC & 21.7 \\ \hline
  \textbf{Intel DevCloud} \cite{Intel-Devcloud} & \scriptsize{\textbf{Intel Xeon E-2176 P630}} & Intel/icpx & 2021.1.2 \\ \hline
  \textbf{Exxact} & \scriptsize{\textbf{NVIDIA GeForce RTX 2080}} & LLVM/clang (nvptx) & 14.0.0 \\ \hline
\end{tabular}
\caption{Cluster and Compilers used in experiment}
\label{tab:advisor:cluster}
\end{table}
In the Kernel Transformation module, we create a vector of seven strings.
The location of these seven strings are shown in Code~\ref{lst:advisor:code_gen}.
The first string (at $\#0$) is always the comment that maintains the text $-$ ``\texttt{Predicted Runtime: \#\# s}''.
As the text suggests \texttt{\#\#} is the predicted runtime for this particular kernel variant.
This string is always placed before the kernel's start location.
Then comes the \texttt{target enter data} construct (at $\#1$).
This directive handles what memory on the GPU needs to be created for the kernel and what data needs to be sent to the GPU before execution.
This string is always placed right after the comments  string.
The next string (at $\#2$) contains the OpenMP directive which specifies that this is the kernel to offload to the target. 
To gain maximum performance out of the GPU, we should always distribute the kernel across all the teams available in the GPU.
Hence this string always contain the directive $-$ ``\texttt{\#pragma omp target teams distribute}''.
The variant determines whether this directive contains any other clauses such as \texttt{collapse} or \texttt{map}, and what will be the values of the clause.
This string is always placed immediately before the kernel's start location, but after the \texttt{target enter data}  string.
The remaining strings ($\#3$, $\#4$ and $\#5$ if required by the variant) are placed just before the start location of their nested \texttt{for} loop.
If these strings are not needed by a variant, they are left empty and no code is inserted in their location.
The last string (at $\#6$) is the \texttt{target exit data} construct, which identifies the data that must be released from the GPU or returned to the CPU.
If not empty, each of these strings is always terminated by a new line.
Once these seven strings are in their proper location, the code is dumped into a new C++ file and returned to the application scientists, who can choose the best code based on the kernel runtime provided in the comment.
They could choose whether to accept or reject the transformations, and then choose which file to build and link with their code.
\section{Experiments and Evaluations}
\label{sec:experiments}

We used several clusters and compilers, as shown in Table~\ref{tab:advisor:cluster}, to test and evaluate our tool.
For the purposes of this study, we only use one GPU per node on the cluster.
The management of multiple GPUs is left for future research.
The three modules explained in Section~\ref{sec:impl} need different experiments.

\begin{figure}[b!]
\vspace{-4mm}
\centering
\pgfplotsset{width=0.95\columnwidth, compat=1.9}

\begin{tikzpicture}
\begin{axis}[xbar=10pt, 
    bar width=4pt,
    xmin=0, 
    xmax=7, 
    bar shift=-2.1pt, 
    yticklabels={,,},
    legend style={at={(axis cs:4.7,0)},anchor=south west, draw=none},
    axis y line*=none,
    axis x line*=bottom,
    xlabel={\textbf{Data Transfer (GB)}},
    ]
\addplot[fill=orange1,postaction={pattern=dots,pattern color=dkred}, draw=orange1] coordinates
{(0.47,0) (0.15,1) (0.30,2) (1.49,3) (1.49,4) (0.30,5) (2.24,6) (0.75,7) (1.49,8) (1.23,9) (3.02,10) (0.75,11)};
\legend{After Transform}
\end{axis}

\begin{axis}[xbar=10pt, 
    bar width=4pt,
    xmin=0, 
    xmax=7, 
    bar shift=2.1pt, 
    yticklabels= {bfs, correlation, covariance, gauss, laplace, knn, mm, mv, mt, particle, proxy\_app, wilson\_dslash},
    ytick={0,1,2,3,4,5,6,7,8,9,10,11},
    legend style={at={(axis cs:4.7,1)},anchor=south west, draw=none},
    axis y line*=none,
    axis x line*=bottom,
    ]
\addplot[fill=dkred,postaction={pattern=dots,pattern color=orange1!30}, draw=dkred] coordinates
{(0.78,0) (0.30,1) (0.60,2) (1.49,3) (2.98,4) (0.45,5) (4.47,6) (1.49,7) (2.98,8) (2.32,9) (6.05,10) (1.50,11)};
\legend{Before Transform}
\end{axis}
\end{tikzpicture}

\caption{Total Data transfer for all Benchmarks Before and After Data Analysis}
\label{plot:data}
\end{figure}
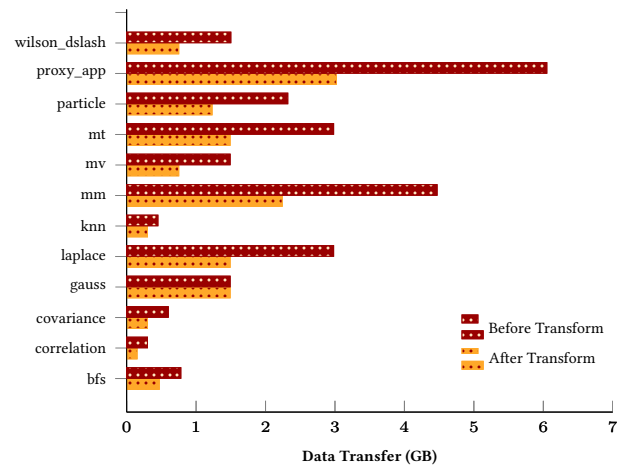
\subsection{Experiment 1 - Data Analysis}
First, we test our Advisor against all benchmark applications to determine whether or not data is correctly identified and generated.
In order to conduct this experiment, we made use of our advisor to generate code that used the correct data between the host and the device.
Additionally, we manually altered each benchmark algorithm to map all data to and from the GPU.
We executed all the codes on each cluster, as shown in Table~\ref{tab:advisor:cluster}, and gathered data about the volume and the duration of the data transfer.
\begin{figure}[t!]
\centering
\pgfplotsset{width=0.95\columnwidth, height=4cm, compat=1.9}

\begin{tikzpicture}
\begin{axis}[ybar=10pt, 
    bar width=8pt,
    xmin=0,
    ymin=0,
    ymax=600,
    bar shift=-4.1pt, 
    xticklabels= {,Summit, Corona, Ookami, Wombat, Seawulf, Intel, Exxact},
    ylabel={\textbf{Data Transfer Time (ms)}},
    xtick={0,1,2,3,4,5,6,7},
    legend style={at={(axis cs:0,610)},anchor=north west, draw=none},
    yticklabels={,,},
    axis y line*=none,
    axis x line*=bottom,
    y label style={at={(axis description cs:-0.07,.5)},anchor=south},
    ]
\addplot[fill=dkred,postaction={pattern=dots,pattern color=orange1}, draw=dkred] coordinates
{(0,0) (1,436.19) (2,214.53) (3,551.70) (4,460.49) (5,520.30) (6,401.33) (7,571.05)};
\legend{Before Transform}
\end{axis}

\begin{axis}[ybar=10pt, 
    bar width=8pt,
    xmin=0,
    ymin=0,
    ymax=600,
    bar shift=4.1pt, 
    xticklabels= {,,},
    xtick={0,1,2,3,4,5,6,7},
    legend style={at={(axis cs:0,510)},anchor=north west, draw=none},
    axis y line*=none,
    axis x line*=bottom,
    ]
\addplot[fill=orange1,postaction={pattern=dots,pattern color=dkred}, draw=orange1] coordinates
{(0,0) (1,328.27) (2,108.85) (3,417.74) (4,318.94) (5,286.67) (6,341.33) (7,419.33)};
\legend{After Transform}
\end{axis}

\end{tikzpicture}
\caption{Data Transfer time (ms) for Wilson Dslash Operator on different Clusters for Before (1.5GB) and After (0.75GB) transformation}
\label{plot:transfertime}
\vspace{-4mm}
\end{figure}
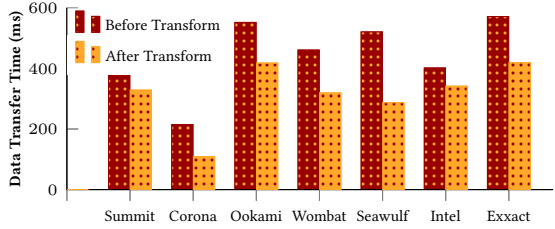
We found that the Advisor improved the data management in all cases.
Figure~\ref{plot:data} shows the amount of data transferred (in GB) between the CPU and the GPU before and after transformation for all benchmark applications.
After applying our transformation, we can clearly see that the amount of data transfer has indeed been considerably reduced.
Reduced data transfer has an impact on all applications' data transfer times.
On all the available clusters, we ran these applications and collected the data transfer times.
Figure~\ref{plot:transfertime} shows the data transfer time for the Wilson Dslash Operator across different clusters.

\subsection{Experiment 2 - Code Generation}
In the first experiment, the code generation for data analysis has already been tested.
But that experiment does not generate different variants of code.
In the second experiment, we use our Advisor to generate every possible code combination using each of the Benchmark applications, as discussed in Section~\ref{subsec:varGen}.
Table~\ref{tab:advisor:result:codegen} shows the result of this experiment, which matches with the expected result in Table~\ref{tab:advisor:variants}.
We compiled all of these codes for various clusters using the compilers shown in Table~\ref{tab:advisor:cluster}.
Some compilers (NVIDIA/nvc on Seawulf and LLVM/Clang 15 on Wombat) do not support dynamic or guided scheduling on a GPU, resulting in compilation failure.
Apart from that, all of the codes successfully compiled and ran on their respective clusters.
We collected the runtime of each of the kernels in this experiment, to be used by our cost model.

We collected the data for the Intel architecture a while ago, and we don't currently have access to the cluster to conduct new experiments.
As a result we had very limited data for Wilson Dslash Operator and no data for our proxy\_app on the Intel architecture.
We were unable to gather many data points for the Exxact machine (with NVIDIA GeForce GPUs) due to the unavailability of compute nodes.
Both these clusters has only around $2,000$ data points each.

\begin{table}[t!]
\centering
\small
\begin{tabular}{|L{0.27\columnwidth}|C{0.18\columnwidth}|C{0.16\columnwidth}|C{0.2\columnwidth}|}
\hline
\rowcolor{black}
\textcolor{white}{\textbf{Application}} & \textcolor{white}{\textbf{Kernel}} & \textcolor{white}{\textbf{\# Loops}} & \textcolor{white}{\textbf{Variants Generated}} \\ \hline
 & kernel1 & 1 & 6 \\ \cline{2-4}
\multirow{-2}{*}{\textbf{bfs}} & kernel2 & 1 & 6 \\ \hline
\textbf{correlation} & kernel1 & 1 & 6 \\ \hline
 & kernel1 & 1 & 6 \\ \cline{2-4}
\multirow{-2}{*}{\textbf{covariance}} & kernel2 & 1 & 6 \\ \hline
\textbf{gauss} & kernel1 & 2 & 18 \\ \hline
& kernel1 & 2 & 18 \\ \cline{2-4}
\multirow{-2}{*}{\textbf{laplace}} & kernel2 & 2 & 18 \\ \hline
\textbf{knn} & kernel1 & 1 & 6 \\ \hline
\textbf{mm} & kernel1 & 2 & 18 \\ \hline
\textbf{mv} & kernel1 & 1 & 6 \\ \hline
\textbf{mt} & kernel1 & 2 & 18 \\ \hline
 & kernel1 & 1 & 6 \\ \cline{2-4}
 & kernel2 & 1 & 6 \\ \cline{2-4}
 & kernel3 & 1 & 6 \\ \cline{2-4}
 & kernel4 & 1 & 6 \\ \cline{2-4}
 & kernel5 & 1 & 6 \\ \cline{2-4}
 & kernel6 & 1 & 6 \\ \cline{2-4}
\multirow{-7}{*}{\textbf{particle}} & kernel7 & 1 & 6 \\ \hline
\textbf{Proxy App} & kernel1 & 4 & 84 \\ \hline
\textbf{Wilson Dslash} & kernel1 & 4 & 84 \\ \hline

\end{tabular}
\caption{Number of variants generated for all applications} 
\label{tab:advisor:result:codegen}
\vspace{-4mm}
\end{table}
Seawulf has NVIDIA K80 GPUs, which is the slowest of the GPUs we're using in our experiment.
So each kernel runs longer on Seawulf than it would on any other cluster.
On the other hand, most variants of kernels failed to compile on Wombat due to their compilers not supporting dynamic and guided scheduling on GPUs.
Due to these reasons, we could only collect around $3,000$ data points on Seawulf and Wombat.
All our kernels compiled and ran successfully on Summit, Corona and Ookami and we were able to collect over $10,000$ data points on each of these architectures.

\subsection{Experiment 3 - Cost Model}
To build our cost model, we extended the COMPOFF model from six variants to all 84 variants.
We build our cost model in the testing mode and then use it to predict the runtime in the prediction mode.
Our cost model utilizes an MLP model with six layers: one input layer, four hidden layers, and one output layer.
We set the number of neurons on multiples of the number of input features rather than choosing a random number of neurons in each hidden layer or conducting an exhaustive grid search (number of neurons in the first layer).
As a result, the first, second, third, and fourth hidden layers, with 33 input features, have 66, 132, 66, and 33 neurons, respectively.

The weights of linear layers are set using the glorot initialization method, which is described in~\cite{glorot2010understanding}.
The bias is set to 0, and the batch size for training data is set to 16 in all runs.
As the underlying optimization algorithm, we evaluate SGD (Stochastic Gradient Descent), Adam~\cite{adamdk} and RMSprop~\cite{hinton2012neural}.
Since it provides optimal performance on transformations for both HPC clusters, we chose the RMSprop optimization algorithm with an initial learning rate of 0.01 that is stepped down by a factor of 0.1 every 30 epochs and weight decay of 0.0001 for 150 epochs.
The Root Mean Square Error~(RMSE) loss function is used to train all models, as defined by Equation~\ref{eq:rmse}, where $\bar{y_i}$ and $y_i$  represent the predicted and ground truth runtimes, respectively.
\begin{equation}\label{eq:rmse}
\begin{aligned}
RMSE(\bar{y}, y) = \sqrt{\frac{1}{N} \sum_{i=0}^{N}(\bar{y_i} - y_i)^2}
\end{aligned}
\end{equation}

\begin{figure}[t!]
    \includegraphics[width=\columnwidth]{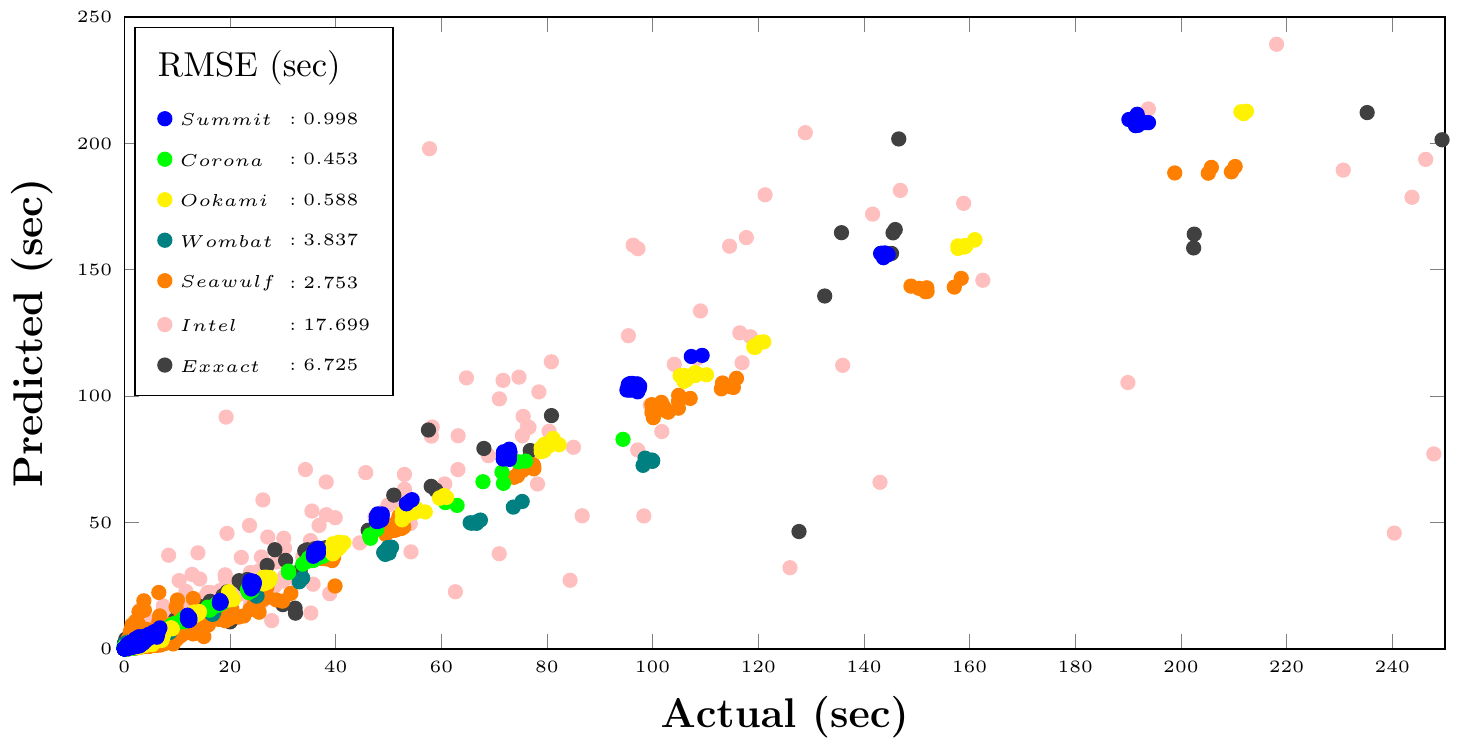}
	\caption{Validation of Cost Model on different clusters}
	\label{plot:validate}
	\vspace{-2mm}
\end{figure}
We split the dataset used by all benchmark applications into two parts: training (80\%) and validation(20\%).
The validation set do not occur in any learning for the model.
The only augmentation applied to the training and validation data is Z-score standardization.
The model is trained using the training set, and after that, testing data are given to the model to test its performance.
In order to determine the standard deviation of the prediction errors, we compute the RMSE. 
The lower this value, the better our model.
However, what constitutes a good RMSE is determined by the range of data on which we are computing RMSE.
One way to determine whether an RMSE value is ``good'' is to normalize it using the following formula:
\vspace{-2mm}
\begin{equation}\label{eq:nrmse}
\begin{aligned}
NRMSE(\bar{y}, y) = \frac{RMSE(\bar{y},y)}{(y_{max} - y_{min})}
\end{aligned}
\vspace{-1mm}
\end{equation}
This yields a value between 0 and 1, with values closer to 0 indicating better fitting models.
Having a model with a normalized RMSE of less than 0.1 is considered successful in this study.

We can see that a simple MLP performs admirably in all of our applications, as evidenced by the strong correlation between actual and predicted data in Figure~\ref{plot:validate}.
It was anticipated that Intel and Exxact's model would perform the worst because of the lack of data.
However, the model for Exxact performed much better (and also showed signs of convergence) than that on Intel because we were able to collect some data for our proxy app on the Exxact system.
Both Wombat and Seawulf performed moderately well when compared to other models trained on a larger dataset.
It is still an open question that how much data is enough data to train an ML model.
We have observed from Figure~\ref{plot:validate}, however, that if we have more than $10,000$ data points for our model, we will be able to train a model that is much more acceptable.

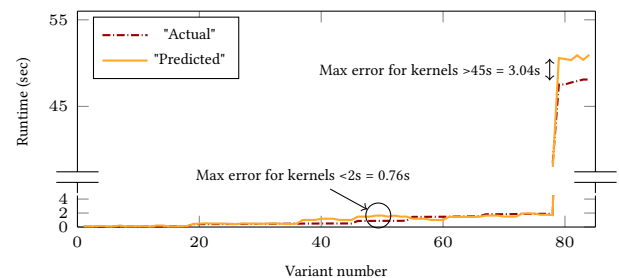
\begin{figure}[b!]
\vspace{-4mm}
\centering
\pgfplotsset{width=\columnwidth, compat=1.9}

\begin{tikzpicture}
\begin{groupplot}[
    group style={
        group size=1 by 2,
        xticklabels at=edge bottom,
        vertical sep=1pt
    },
    xmin=0, xmax=85,
]

\nextgroupplot[ymin=35,ymax=56,
               ytick={45,55},
               ylabel={Runtime (sec)},
               axis x line=top, 
               axis y discontinuity=parallel,
               height=4cm,
               legend pos=north west]
               
\addplot[dash pattern=on 3pt off 1pt on \the\pgflinewidth off 1pt, thick, dkred] coordinates{
(78,38.5)
(79,47.532772)
(80,47.533264)
(81,47.753544)
(82,47.922302)
(83,48.098248)
(84,48.107327)
};

\addplot[thick, orange1] coordinates{
(78,38)
(79,50.575325)
(80,50.47608)
(81,50.34886)
(82,50.900986)
(83,50.38127)
(84,50.957832)
};

\draw (775,130) edge[<->] node[left] {\tiny{Max error for kernels >45s = 3.04s}} (775,155);

\legend{"Actual","Predicted"};

\nextgroupplot[ymin=0,ymax=4.5,
               ytick={0,2,4},
               axis x line=bottom,
               xtick={0,20,40,60,80},
               xlabel={Variant number},
               height=2.0cm]

\addplot[dash pattern=on 3pt off 1pt on \the\pgflinewidth off 1pt, thick, dkred] coordinates{
(1,0.041991)
(2,0.043546)
(3,0.046941)
(4,0.04729)
(5,0.052052)
(6,0.05257)
(7,0.059971)
(8,0.063517)
(9,0.066341)
(10,0.067006)
(11,0.067393)
(12,0.06746)
(13,0.085273)
(14,0.085295)
(15,0.096467)
(16,0.096965)
(17,0.098902)
(18,0.099118)
(19,0.408377)
(20,0.414542)
(21,0.417347)
(22,0.418222)
(23,0.423778)
(24,0.427283)
(25,0.433373)
(26,0.434397)
(27,0.43534)
(28,0.43673)
(29,0.441469)
(30,0.448989)
(31,0.458964)
(32,0.459303)
(33,0.468977)
(34,0.473851)
(35,0.476765)
(36,0.476848)
(37,0.492259)
(38,0.496883)
(39,0.497887)
(40,0.502192)
(41,0.507599)
(42,0.508104)
(43,0.509212)
(44,0.510813)
(45,0.519478)
(46,0.866311)
(47,0.871272)
(48,0.874585)
(49,0.877332)
(50,0.884297)
(51,0.884409)
(52,0.884704)
(53,0.886242)
(54,0.907998)
(55,1.453897)
(56,1.456185)
(57,1.457746)
(58,1.464329)
(59,1.46551)
(60,1.46866)
(61,1.470907)
(62,1.500985)
(63,1.505508)
(64,1.52314)
(65,1.530806)
(66,1.532966)
(67,1.827711)
(68,1.828468)
(69,1.836358)
(70,1.839236)
(71,1.83969)
(72,1.866009)
(73,1.872533)
(74,1.876093)
(75,1.878451)
(76,1.890767)
(77,1.891543)
(78,1.896286)
(79,47.532772)
};

\addplot[thick, orange1] coordinates{
(1,0.08900112)
(2,0.07211521)
(3,0.14284557)
(4,0.14888787)
(5,0.08900112)
(6,0.08900112)
(7,0.21584825)
(8,0.08900112)
(9,0.09187369)
(10,0.12973398)
(11,0.10806763)
(12,0.07784614)
(13,0.21584825)
(14,0.21584825)
(15,0.08900112)
(16,0.08900112)
(17,0.12973398)
(18,0.12973398)
(19,0.45056686)
(20,0.48421472)
(21,0.5299194)
(22,0.50127244)
(23,0.50127244)
(24,0.50127244)
(25,0.4462544)
(26,0.40385666)
(27,0.505718)
(28,0.48650536)
(29,0.50127244)
(30,0.4277597)
(31,0.505718)
(32,0.505718)
(33,0.50127244)
(34,0.4277597)
(35,0.4277597)
(36,0.50127244)
(37,1.0175349)
(38,1.0175349)
(39,1.0175349)
(40,1.1876421)
(41,1.1876421)
(42,1.1876421)
(43,1.0175349)
(44,1.0175349)
(45,1.0175349)
(46,1.489763)
(47,1.489763)
(48,1.489763)
(49,1.6390233)
(50,1.6390233)
(51,1.489763)
(52,1.6390233)
(53,1.489763)
(54,1.489763)
(55,1.1876421)
(56,1.1876421)
(57,1.1876421)
(58,1.0175349)
(59,1.0175349)
(60,1.0175349)
(61,1.4561753)
(62,1.4561753)
(63,1.4561753)
(64,1.4522922)
(65,1.4618161)
(66,1.4794965)
(67,1.6390233)
(68,1.6390233)
(69,1.6390233)
(70,1.489763)
(71,1.489763)
(72,1.489763)
(73,1.9416566)
(74,1.9416566)
(75,1.9416566)
(76,1.7454329)
(77,1.7633965)
(78,1.7104163)
(79,51.575325)
};

\end{groupplot}

\node(a) at (3,0.25) {\tiny{Max error for kernels <2s = 0.76s}};
\node[circle,draw](c) at (4.0,-0.3){};
\draw[->] (a) -- (c);

\end{tikzpicture}
\caption{Wilson Dslash operator's Actual and Predicted Runtimes (in sec) on Summit for all variants sorted by runtime, where $L_X, L_Y, L_Z$ and $L_T$ are set to $32, 32, 32$ and $16$.}

\label{plot:plot_variants_summit}
\end{figure}

\subsection{Experiment 4 - Prediction}
\begin{figure}[t!]
\vspace{-2mm}
\centering
\pgfplotsset{width=0.95\columnwidth, compat=1.9}

\begin{tikzpicture}
\begin{groupplot}[
    group style={
        group size=1 by 2,
        xticklabels at=edge bottom,
        vertical sep=2pt
    },
    xmin=0, xmax=7.2,
]

\nextgroupplot[
    ybar=10pt, 
    bar width=8pt,
    xmin=0,
    xmax=7.2,
    ymin=0,
    ymax=0.5,
    ylabel={\textbf{Normalized RMSE}},
    ylabel style={rotate=-90,align=right,text width=1cm},
    xtick={0,1,2,3,4,5,6,7},
    xticklabel style={align=center,text width=1cm},
    axis y line*=none,
    axis x line=top, 
    height=2.5cm,
    y dir=reverse,
    ]
\addplot[fill=dkred,postaction={pattern=dots,pattern color=orange1}, draw=dkred] coordinates
{ (1,0.02) (2,0.008) (3,0.008) (4,0.033) (5,0.066) (6,0) (7,0.279) };

\nextgroupplot[
    ybar=10pt, 
    bar width=8pt,
    xmin=0,
    xmax=7.2,
    ymin=0,
    ylabel={\textbf{RMSE(s)}},
    ylabel style={rotate=-90,align=right,text width=1cm},
    xtick={0,1,2,3,4,5,6,7},
    xticklabels= {,Summit [0 - 49s], Corona [0 - 23s], Ookami [0 - 53s], Wombat [0 - 65s], Seawulf [0 - 103s], Intel, Exxact [0 - 40s]},
    ytick={3,6,9},
    xticklabel style={align=center,text width=1cm},
    axis y line*=none,
    axis x line*=bottom,
    height=3cm,
    ]
\addplot[fill=dkred,postaction={pattern=dots,pattern color=orange1}, draw=dkred] coordinates
{ (1,0.998) (2,0.190) (3,0.447) (4,4.273) (5,3.375) (6,0) (7,11.153) };

\end{groupplot}

\node(summit2) at (0.9,0.7) {(\textcolor{red}{0.020})};
\node(corona2) at (1.8,0.75) {(\textcolor{red}{0.008})};
\node(ookami2) at (2.7,0.75) {(\textcolor{red}{0.008})};
\node(wombat2) at (3.6,0.65) {(\textcolor{red}{0.033})};
\node(seawulf2) at (4.5,0.6) {(\textcolor{red}{0.066})};
\node(exxact2) at (5.4,0.75) {(\textcolor{red}{\texttt{-NA-}})};
\node(intel2) at (6.3,0.25) {(\textcolor{red}{0.279})};

\node(summit) at (0.9,-1.15) {0.998};
\node(corona) at (1.8,-1.25) {0.190};
\node(ookami) at (2.7,-1.25) {0.447};
\node(wombat) at (3.6,-0.8) {4.273};
\node(seawulf) at (4.5,-0.9) {3.375};
\node(intel) at (5.4,-1.3) {\texttt{-NA-}};
\node(exxact) at (6.3,-0.05) {11.153};
\end{tikzpicture}
\caption{RMSE and Normalized RMSE for predicting the runtime of all variants of Wilson Dslash operator on different clusters for $L_X,L_Y,L_Z,L_T$ set at $32,32,32,16$ each. The range of runtimes for each cluster is mentioned below their name.}
\label{plot:plotRMSE}
\end{figure}
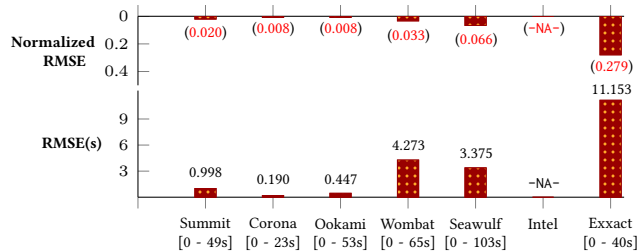
Finally, for our final set of experiments we use our Advisor to predict the top 10 best variants for the Wilson Dslash Operator.
Once the top 10 variants are identified we use the Code Transformation module to generate those 10 code variants and return them back to the user.
The Advisor takes as input the base Wilson Dslash kernel, and generates the top 10 best kernels as predicted by the cost model.
As shown in Figure~\ref{plot:plot_variants_summit}, we plot the actual and predicted runtimes of all the 84 generated variants (sorted by actual runtime) of one such kernel when run on the Summit supercomputer.
We can clearly see a strong correlation between the actual and predicted runtime for all the variants.
The same correlation can be found in almost all kernels across all clusters.
In Figure~\ref{plot:plotRMSE}, we display the Wilson Dslash operator's RMSE and normalized RMSE for each cluster.
The range of runtimes (in seconds) for each cluster is mentioned below their name in the plot.
We currently do not have access to the Intel cluster to conduct new experiments, and the Intel dataset contained very few data from the targeted Wilson Dslash kernel and none from our proxy\_app.
So, even if we make a prediction using this model, there is no way to validate it.
Consequently, we did not conduct this experiment on the Intel architecture and the result is marked as $-NA-$.
As expected on Exxact, the target kernel's RMSE increased significantly (11.153s) due to less data in its dataset.
Even with a normalized RMSE of 0.279, it fell short of our expectation of 0.1.
Nonetheless, this model demonstrated some correlation between the actual and predicted data.
In contrast, Wombat and Seawulf performed reasonably well and were able to predict the top 10 kernel variants despite having an RMSE of 4.273s and 3.375s, respectively.
However, with 0.033 and 0.066, respectively, their normalized RMSE was well within our expectation.
As per our observation, their RMSE can also be improved by adding more data for these clusters.
Finally, as shown in Figure~\ref{plot:plotRMSE}, the RMSE rates for Summit, Corona, and Ookami are less than one second each, and they were able to accurately predict the top ten kernel variants.
\section{Conclusion and Future Work}
\label{sec:conclusion}

In this paper, we introduced the OpenMP Advisor, a compiler tool that advises application scientists on various OpenMP code offloading strategies.
Although the tool is currently restricted to GPUs, it can be extended to support other OpenMP-capable devices.
Using our Advisor, we were able to generate code of multiple applications for seven different architectures, and correctly predict the top ten best variants for each application including a real world application (the Wilson Dslash operator) on every architecture.
Preliminary findings indicate that this tool can assist compiler developers and HPC application scientists in porting their legacy HPC codes to the upcoming heterogeneous computing environment.
As a next step, we will extend our tool to include the following tasks:
\begin{itemize}[leftmargin=*]
    \item Data synchronization between host and device during kernel execution.
    \item Offload computation to multiple GPUs~\cite{kale2020toward} via tasks.
    \item Expand benchmarks to include more kernels to cover a wider range of applications.
    \item Collect more data for all architectures and expand our research to include new GPUs and compilers.
    \item Our cost model can currently predict the runtime of kernels whose data size and level of parallelism are known at compile time. 
    We could extend our model to predict the best variants for a variety of data sizes, and then use the OpenMP metadirective~\cite{mishra2020extending} directive to generate multiple directive variants for each range.
    \item According to previous studies \cite{mishra2017benchmarking,li2018manage}, OpenMP GPU offloading performance can be enhanced by using unified memory between the CPU and GPU and such analysis should be a part of our Advisor.
    \item Support OpenMP Advisor for all OpenMP directives, not just target offloading.
    \item Extend the Advisor to other directive-based models, such as OpenACC.
\end{itemize}

This tool is a first-of-its-kind attempt to build a framework for automatic GPU offloading using OpenMP and machine learning; as a result, there is plenty of room for improvement.
\section*{Acknowledgement}
This research was supported by the Exascale Computing Project (17-SC-20-SC), a collaborative effort of the U.S. Department of Energy Office of Science and the National Nuclear Security Administration. 
This material is also based upon work supported by the National Science Foundation under grant no. CCF-2113996.
This research used resources of the Oak Ridge Leadership Computing Facility at the Oak Ridge National Laboratory, which is supported by the Office of Science of the U.S. Department of Energy under Contract No. DE-AC05-00OR22725.
The authors would like to thank Stony Brook Research Computing and Cyberinfrastructure, and the Institute for Advanced Computational Science at Stony Brook University for access to the SeaWulf computing system, which was made possible by  a \$1.4M National Science Foundation grant (\#1531492).

\bibliography{main}


\begin{thebibliography}{40}


\ifx \showCODEN    \undefined \def \showCODEN     #1{\unskip}     \fi
\ifx \showDOI      \undefined \def \showDOI       #1{#1}\fi
\ifx \showISBNx    \undefined \def \showISBNx     #1{\unskip}     \fi
\ifx \showISBNxiii \undefined \def \showISBNxiii  #1{\unskip}     \fi
\ifx \showISSN     \undefined \def \showISSN      #1{\unskip}     \fi
\ifx \showLCCN     \undefined \def \showLCCN      #1{\unskip}     \fi
\ifx \shownote     \undefined \def \shownote      #1{#1}          \fi
\ifx \showarticletitle \undefined \def \showarticletitle #1{#1}   \fi
\ifx \showURL      \undefined \def \showURL       {\relax}        \fi
\providecommand\bibfield[2]{#2}
\providecommand\bibinfo[2]{#2}
\providecommand\natexlab[1]{#1}
\providecommand\showeprint[2][]{arXiv:#2}

\bibitem[Barua et~al\mbox{.}(2019)]%
        {barua2019ompsan}
\bibfield{author}{\bibinfo{person}{Prithayan Barua}, \bibinfo{person}{Jun
  Shirako}, \bibinfo{person}{Whitney Tsang}, \bibinfo{person}{Jeeva Paudel},
  \bibinfo{person}{Wang Chen}, {and} \bibinfo{person}{Vivek Sarkar}.}
  \bibinfo{year}{2019}\natexlab{}.
\newblock \showarticletitle{OMPSan: static verification of openmp’s data
  mapping constructs}. In \bibinfo{booktitle}{\emph{International Workshop on
  OpenMP}}. Springer, \bibinfo{address}{Auckland, New Zealand},
  \bibinfo{pages}{3--18}.
\newblock


\bibitem[{Brower, Richard} et~al\mbox{.}(2018)]%
        {refId0}
\bibfield{author}{\bibinfo{person}{{Brower, Richard}},
  \bibinfo{person}{{Christ, Norman}}, \bibinfo{person}{{DeTar, Carleton}},
  \bibinfo{person}{{Edwards, Robert}}, {and} \bibinfo{person}{{Mackenzie,
  Paul}}.} \bibinfo{year}{2018}\natexlab{}.
\newblock \showarticletitle{Lattice QCD Application Development within the US
  DOE Exascale Computing Project}.
\newblock \bibinfo{journal}{\emph{EPJ Web Conf.}}  \bibinfo{volume}{175}
  (\bibinfo{year}{2018}), \bibinfo{pages}{09010}.
\newblock
\urldef\tempurl%
\url{https://doi.org/10.1051/epjconf/201817509010}
\showDOI{\tempurl}


\bibitem[Burford et~al\mbox{.}(2021)]%
        {burford2021ookami}
\bibfield{author}{\bibinfo{person}{Andrew Burford}, \bibinfo{person}{Alan
  Calder}, \bibinfo{person}{David Carlson}, \bibinfo{person}{Barbara Chapman},
  \bibinfo{person}{Firat Coskun}, \bibinfo{person}{Tony Curtis},
  \bibinfo{person}{Catherine Feldman}, \bibinfo{person}{Robert Harrison},
  \bibinfo{person}{Yan Kang}, \bibinfo{person}{Benjamin Michalowicz},
  {et~al\mbox{.}}} \bibinfo{year}{2021}\natexlab{}.
\newblock \showarticletitle{Ookami: Deployment and Initial Experiences}.
\newblock In \bibinfo{booktitle}{\emph{Practice and Experience in Advanced
  Research Computing}}. \bibinfo{publisher}{ACM}, \bibinfo{address}{New York},
  \bibinfo{pages}{1--8}.
\newblock


\bibitem[Che et~al\mbox{.}(2009)]%
        {che2009rodinia}
\bibfield{author}{\bibinfo{person}{Shuai Che}, \bibinfo{person}{Michael Boyer},
  \bibinfo{person}{Jiayuan Meng}, \bibinfo{person}{David Tarjan},
  \bibinfo{person}{Jeremy~W Sheaffer}, \bibinfo{person}{Sang-Ha Lee}, {and}
  \bibinfo{person}{Kevin Skadron}.} \bibinfo{year}{2009}\natexlab{}.
\newblock \showarticletitle{Rodinia: A benchmark suite for heterogeneous
  computing}. In \bibinfo{booktitle}{\emph{2009 IEEE international symposium on
  workload characterization (IISWC)}}. Ieee, \bibinfo{pages}{44--54}.
\newblock


\bibitem[Clang, Plugins(2022)]%
        {clangplugin}
Clang, Plugins \bibinfo{year}{2022}\natexlab{}.
\newblock \bibinfo{title}{Clang Plugins}.
\newblock
\newblock
\urldef\tempurl%
\url{https://clang.llvm.org/docs/ClangPlugins.html}
\showURL{%
Retrieved Apr 20, 2022 from \tempurl}


\bibitem[Dagum and Menon(1998)]%
        {dagum1998openmp}
\bibfield{author}{\bibinfo{person}{Leonardo Dagum} {and}
  \bibinfo{person}{Ramesh Menon}.} \bibinfo{year}{1998}\natexlab{}.
\newblock \showarticletitle{OpenMP: an industry standard API for shared-memory
  programming}.
\newblock \bibinfo{journal}{\emph{IEEE computational science and engineering}}
  \bibinfo{volume}{5}, \bibinfo{number}{1} (\bibinfo{year}{1998}),
  \bibinfo{pages}{46--55}.
\newblock


\bibitem[Dennard et~al\mbox{.}(1974)]%
        {dennard1974design}
\bibfield{author}{\bibinfo{person}{Robert~H Dennard}, \bibinfo{person}{Fritz~H
  Gaensslen}, \bibinfo{person}{Hwa-Nien Yu}, \bibinfo{person}{V~Leo Rideout},
  \bibinfo{person}{Ernest Bassous}, {and} \bibinfo{person}{Andre~R LeBlanc}.}
  \bibinfo{year}{1974}\natexlab{}.
\newblock \showarticletitle{Design of ion-implanted MOSFET's with very small
  physical dimensions}.
\newblock \bibinfo{journal}{\emph{IEEE Journal of solid-state circuits}}
  \bibinfo{volume}{9}, \bibinfo{number}{5} (\bibinfo{year}{1974}),
  \bibinfo{pages}{256--268}.
\newblock


\bibitem[Falkner et~al\mbox{.}(2018)]%
        {falkner2018bohb}
\bibfield{author}{\bibinfo{person}{Stefan Falkner}, \bibinfo{person}{Aaron
  Klein}, {and} \bibinfo{person}{Frank Hutter}.}
  \bibinfo{year}{2018}\natexlab{}.
\newblock \showarticletitle{BOHB: Robust and efficient hyperparameter
  optimization at scale}. In \bibinfo{booktitle}{\emph{International Conference
  on Machine Learning}}. PMLR, \bibinfo{pages}{1437--1446}.
\newblock


\bibitem[Gelado et~al\mbox{.}(2010)]%
        {gelado2010asymmetric}
\bibfield{author}{\bibinfo{person}{Isaac Gelado}, \bibinfo{person}{John~E
  Stone}, \bibinfo{person}{Javier Cabezas}, \bibinfo{person}{Sanjay Patel},
  \bibinfo{person}{Nacho Navarro}, {and} \bibinfo{person}{Wen-mei~W Hwu}.}
  \bibinfo{year}{2010}\natexlab{}.
\newblock \showarticletitle{An asymmetric distributed shared memory model for
  heterogeneous parallel systems}. In \bibinfo{booktitle}{\emph{Proceedings of
  the fifteenth International Conference on Architectural support for
  programming languages and operating systems}}. \bibinfo{pages}{347--358}.
\newblock


\bibitem[Glorot and Bengio(2010)]%
        {glorot2010understanding}
\bibfield{author}{\bibinfo{person}{Xavier Glorot} {and} \bibinfo{person}{Yoshua
  Bengio}.} \bibinfo{year}{2010}\natexlab{}.
\newblock \showarticletitle{Understanding the difficulty of training deep
  feedforward neural networks}. In \bibinfo{booktitle}{\emph{Proceedings of the
  thirteenth international conference on artificial intelligence and
  statistics}}. JMLR Workshop and Conference Proceedings,
  \bibinfo{pages}{249--256}.
\newblock


\bibitem[Hinton et~al\mbox{.}(2012)]%
        {hinton2012neural}
\bibfield{author}{\bibinfo{person}{Geoffrey Hinton}, \bibinfo{person}{Nitish
  Srivastava}, {and} \bibinfo{person}{Kevin Swersky}.}
  \bibinfo{year}{2012}\natexlab{}.
\newblock \showarticletitle{Neural networks for machine learning lecture 6a
  overview of mini-batch gradient descent}.
\newblock \bibinfo{journal}{\emph{Cited on}} \bibinfo{volume}{14},
  \bibinfo{number}{8} (\bibinfo{year}{2012}), \bibinfo{pages}{2}.
\newblock


\bibitem[Intel(2021)]%
        {Intel-Devcloud}
\bibfield{author}{\bibinfo{person}{Intel}.} \bibinfo{year}{2021}\natexlab{}.
\newblock \bibinfo{title}{Intel {D}eveloper {C}loud}.
\newblock
\newblock
\urldef\tempurl%
\url{https://www.intel.com/content/www/us/en/developer/tools/devcloud/overview.html}
\showURL{%
Retrieved October 25, 2022 from \tempurl}


\bibitem[Jablin et~al\mbox{.}(2011)]%
        {jablin2011automatic}
\bibfield{author}{\bibinfo{person}{Thomas~B Jablin}, \bibinfo{person}{Prakash
  Prabhu}, \bibinfo{person}{James~A Jablin}, \bibinfo{person}{Nick~P Johnson},
  \bibinfo{person}{Stephen~R Beard}, {and} \bibinfo{person}{David~I August}.}
  \bibinfo{year}{2011}\natexlab{}.
\newblock \showarticletitle{Automatic CPU-GPU communication management and
  optimization}. In \bibinfo{booktitle}{\emph{Proceedings of the 32nd ACM
  SIGPLAN conference on Programming language design and implementation}}.
  \bibinfo{pages}{142--151}.
\newblock


\bibitem[Kale et~al\mbox{.}(2020)]%
        {kale2020toward}
\bibfield{author}{\bibinfo{person}{Vivek Kale}, \bibinfo{person}{Wenbin Lu},
  \bibinfo{person}{Anthony Curtis}, \bibinfo{person}{Abid~M Malik},
  \bibinfo{person}{Barbara Chapman}, {and} \bibinfo{person}{Oscar Hernandez}.}
  \bibinfo{year}{2020}\natexlab{}.
\newblock \showarticletitle{Toward supporting multi-GPU targets via taskloop
  and user-defined schedules}. In \bibinfo{booktitle}{\emph{International
  Workshop on OpenMP}}. Springer, \bibinfo{pages}{295--309}.
\newblock


\bibitem[Kedia(2009)]%
        {kedia2009hybrid}
\bibfield{author}{\bibinfo{person}{Kushal Kedia}.}
  \bibinfo{year}{2009}\natexlab{}.
\newblock \bibinfo{booktitle}{\emph{Hybrid programming with openmp and mpi}}.
\newblock \bibinfo{type}{{T}echnical {R}eport}. \bibinfo{institution}{Technical
  Report 18.337 J, Massachusetts Institute of Technology}.
\newblock


\bibitem[Kingma and Ba(2015)]%
        {adamdk}
\bibfield{author}{\bibinfo{person}{Diederik~P. Kingma} {and}
  \bibinfo{person}{Jimmy Ba}.} \bibinfo{year}{2015}\natexlab{}.
\newblock \showarticletitle{Adam: {A} Method for Stochastic Optimization}. In
  \bibinfo{booktitle}{\emph{3rd International Conference on Learning
  Representations, {ICLR} 2015, San Diego, CA, USA, May 7-9, 2015, Conference
  Track Proceedings}}, \bibfield{editor}{\bibinfo{person}{Yoshua Bengio} {and}
  \bibinfo{person}{Yann LeCun}} (Eds.).
\newblock
\urldef\tempurl%
\url{http://arxiv.org/abs/1412.6980}
\showURL{%
\tempurl}


\bibitem[Laboratory(2019)]%
        {Corona}
\bibfield{author}{\bibinfo{person}{Lawrence Livermore~National Laboratory}.}
  \bibinfo{year}{2019}\natexlab{}.
\newblock \bibinfo{title}{{LLNL} - {C}orona}.
\newblock
\newblock
\urldef\tempurl%
\url{https://hpc.llnl.gov/hardware/compute-platforms/corona}
\showURL{%
Retrieved October 25, 2022 from \tempurl}


\bibitem[Li et~al\mbox{.}(2018)]%
        {li2018manage}
\bibfield{author}{\bibinfo{person}{Lingda Li}, \bibinfo{person}{Hal Finkel},
  \bibinfo{person}{Martin Kong}, {and} \bibinfo{person}{Barbara Chapman}.}
  \bibinfo{year}{2018}\natexlab{}.
\newblock \showarticletitle{Manage openmp GPU data environment under unified
  address space}. In \bibinfo{booktitle}{\emph{International Workshop on
  OpenMP}}. Springer, \bibinfo{pages}{69--81}.
\newblock


\bibitem[Lin(2016)]%
        {lin2016optimization}
\bibfield{author}{\bibinfo{person}{M Lin}.} \bibinfo{year}{2016}\natexlab{}.
\newblock \showarticletitle{Optimization of the Domain Wall Dslash Kernel in
  Columbia Physics System}.
\newblock  (\bibinfo{year}{2016}), \bibinfo{pages}{269}.
\newblock


\bibitem[Liu et~al\mbox{.}(2021)]%
        {liu2021gptune}
\bibfield{author}{\bibinfo{person}{Yang Liu}, \bibinfo{person}{Wissam~M
  Sid-Lakhdar}, \bibinfo{person}{Osni Marques}, \bibinfo{person}{Xinran Zhu},
  \bibinfo{person}{Chang Meng}, \bibinfo{person}{James~W Demmel}, {and}
  \bibinfo{person}{Xiaoye~S Li}.} \bibinfo{year}{2021}\natexlab{}.
\newblock \showarticletitle{GPTune: multitask learning for autotuning exascale
  applications}. In \bibinfo{booktitle}{\emph{Proceedings of the 26th ACM
  SIGPLAN Symposium on Principles and Practice of Parallel Programming}}.
  \bibinfo{pages}{234--246}.
\newblock


\bibitem[LLVM(2021)]%
        {llvmcostmodel}
\bibfield{author}{\bibinfo{person}{LLVM}.} \bibinfo{year}{2021}\natexlab{}.
\newblock \bibinfo{booktitle}{\emph{{LLVM} {C}ost {M}odel}}.
\newblock
\urldef\tempurl%
\url{https://llvm.org/doxygen/CostModel\_8cpp.html}
\showURL{%
\tempurl}


\bibitem[Mendon{\c{c}}a et~al\mbox{.}(2017)]%
        {mendoncca2017dawncc}
\bibfield{author}{\bibinfo{person}{Gleison Mendon{\c{c}}a},
  \bibinfo{person}{Breno Guimar{\~a}es}, \bibinfo{person}{P{\'e}ricles Alves},
  \bibinfo{person}{M{\'a}rcio Pereira}, \bibinfo{person}{Guido Ara{\'u}jo},
  {and} \bibinfo{person}{Fernando Magno~Quint{\~a}o Pereira}.}
  \bibinfo{year}{2017}\natexlab{}.
\newblock \showarticletitle{DawnCC: automatic annotation for data parallelism
  and offloading}.
\newblock \bibinfo{journal}{\emph{ACM Transactions on Architecture and Code
  Optimization (TACO)}} \bibinfo{volume}{14}, \bibinfo{number}{2}
  (\bibinfo{year}{2017}), \bibinfo{pages}{13}.
\newblock


\bibitem[Menon et~al\mbox{.}(2020)]%
        {menon2020auto}
\bibfield{author}{\bibinfo{person}{Harshitha Menon}, \bibinfo{person}{Abhinav
  Bhatele}, {and} \bibinfo{person}{Todd Gamblin}.}
  \bibinfo{year}{2020}\natexlab{}.
\newblock \showarticletitle{Auto-tuning parameter choices in HPC applications
  using Bayesian optimization}. In \bibinfo{booktitle}{\emph{2020 IEEE
  International Parallel and Distributed Processing Symposium (IPDPS)}}. IEEE,
  \bibinfo{pages}{831--840}.
\newblock


\bibitem[Mishra et~al\mbox{.}(2022)]%
        {mishra2022compoff}
\bibfield{author}{\bibinfo{person}{Alok Mishra}, \bibinfo{person}{Smeet
  Chheda}, \bibinfo{person}{Carlos Soto}, \bibinfo{person}{Abid~M. Malik},
  \bibinfo{person}{Meifeng Lin}, {and} \bibinfo{person}{Barbara Chapman}.}
  \bibinfo{year}{2022}\natexlab{}.
\newblock \showarticletitle{{COMPOFF}: A Compiler Cost model using Machine
  Learning to predict the Cost of OpenMP Offloading}. In
  \bibinfo{booktitle}{\emph{2022 IEEE International Parallel and Distributed
  Processing Symposium Workshops (IPDPSW), May 30-June 3, 2022}}. IEEE.
\newblock


\bibitem[Mishra et~al\mbox{.}(2017)]%
        {mishra2017benchmarking}
\bibfield{author}{\bibinfo{person}{Alok Mishra}, \bibinfo{person}{Lingda Li},
  \bibinfo{person}{Martin Kong}, \bibinfo{person}{Hal Finkel}, {and}
  \bibinfo{person}{Barbara Chapman}.} \bibinfo{year}{2017}\natexlab{}.
\newblock \showarticletitle{Benchmarking and evaluating unified memory for
  OpenMP GPU offloading}. In \bibinfo{booktitle}{\emph{Proceedings of the
  Fourth Workshop on the LLVM Compiler Infrastructure in HPC}}.
  \bibinfo{pages}{1--10}.
\newblock


\bibitem[Mishra et~al\mbox{.}(2020a)]%
        {mishra2020data}
\bibfield{author}{\bibinfo{person}{Alok Mishra}, \bibinfo{person}{Abid~M
  Malik}, {and} \bibinfo{person}{Barbara Chapman}.}
  \bibinfo{year}{2020}\natexlab{a}.
\newblock \showarticletitle{Data Transfer and Reuse Analysis Tool for
  GPU-Offloading Using OpenMP}. In \bibinfo{booktitle}{\emph{International
  Workshop on OpenMP}}. Springer, \bibinfo{pages}{280--294}.
\newblock


\bibitem[Mishra et~al\mbox{.}(2020b)]%
        {mishra2020extending}
\bibfield{author}{\bibinfo{person}{Alok Mishra}, \bibinfo{person}{Abid~M
  Malik}, {and} \bibinfo{person}{Barbara Chapman}.}
  \bibinfo{year}{2020}\natexlab{b}.
\newblock \showarticletitle{Extending the LLVM/Clang Framework for OpenMP
  Metadirective Support}. In \bibinfo{booktitle}{\emph{2020 IEEE/ACM 6th
  Workshop on the LLVM Compiler Infrastructure in HPC (LLVM-HPC) and Workshop
  on Hierarchical Parallelism for Exascale Computing (HiPar)}}. IEEE,
  \bibinfo{pages}{33--44}.
\newblock


\bibitem[ORNL(2017)]%
        {Summit}
\bibfield{author}{\bibinfo{person}{ORNL}.} \bibinfo{year}{2017}\natexlab{}.
\newblock \bibinfo{title}{{O}ak {R}idge {L}eadership {C}omputing {F}acility -
  {S}ummit Supercomputing cluster}.
\newblock
\newblock
\urldef\tempurl%
\url{https://www.olcf.ornl.gov/summit/}
\showURL{%
Retrieved October 25, 2022 from \tempurl}


\bibitem[ORNL(2020)]%
        {Wombat}
\bibfield{author}{\bibinfo{person}{ORNL}.} \bibinfo{year}{2020}\natexlab{}.
\newblock \bibinfo{title}{{O}ak {R}idge {L}eadership {C}omputing {F}acility -
  {W}ombat cluster}.
\newblock
\newblock
\urldef\tempurl%
\url{https://www.olcf.ornl.gov/olcf-resources/compute-systems/wombat/}
\showURL{%
Retrieved October 25, 2022 from \tempurl}


\bibitem[Poesia et~al\mbox{.}(2017)]%
        {poesia2017static}
\bibfield{author}{\bibinfo{person}{Gabriel Poesia}, \bibinfo{person}{Breno
  Guimar{\~a}es}, \bibinfo{person}{Fabricio Ferracioli}, {and}
  \bibinfo{person}{Fernando Magno~Quint{\~a}o Pereira}.}
  \bibinfo{year}{2017}\natexlab{}.
\newblock \showarticletitle{Static placement of computation on heterogeneous
  devices}.
\newblock \bibinfo{journal}{\emph{Proceedings of the ACM on Programming
  Languages}} \bibinfo{volume}{1}, \bibinfo{number}{OOPSLA}
  (\bibinfo{year}{2017}), \bibinfo{pages}{1--28}.
\newblock


\bibitem[Roy et~al\mbox{.}(2021)]%
        {roy2021bliss}
\bibfield{author}{\bibinfo{person}{Rohan~Basu Roy}, \bibinfo{person}{Tirthak
  Patel}, \bibinfo{person}{Vijay Gadepally}, {and} \bibinfo{person}{Devesh
  Tiwari}.} \bibinfo{year}{2021}\natexlab{}.
\newblock \showarticletitle{Bliss: auto-tuning complex applications using a
  pool of diverse lightweight learning models}. In
  \bibinfo{booktitle}{\emph{Proceedings of the 42nd ACM SIGPLAN International
  Conference on Programming Language Design and Implementation}}.
  \bibinfo{pages}{1280--1295}.
\newblock


\bibitem[Schober et~al\mbox{.}(2018)]%
        {schober2018correlation}
\bibfield{author}{\bibinfo{person}{Patrick Schober}, \bibinfo{person}{Christa
  Boer}, {and} \bibinfo{person}{Lothar~A Schwarte}.}
  \bibinfo{year}{2018}\natexlab{}.
\newblock \showarticletitle{Correlation coefficients: appropriate use and
  interpretation}.
\newblock \bibinfo{journal}{\emph{Anesthesia \& Analgesia}}
  \bibinfo{volume}{126}, \bibinfo{number}{5} (\bibinfo{year}{2018}),
  \bibinfo{pages}{1763--1768}.
\newblock


\bibitem[Suite(2018a)]%
        {gauss-rodinia}
\bibfield{author}{\bibinfo{person}{Rodinia~Benchmark Suite}.}
  \bibinfo{year}{2018}\natexlab{a}.
\newblock \bibinfo{booktitle}{\emph{Gaussian Elimination}}.
\newblock
\urldef\tempurl%
\url{https://www.cs.virginia.edu/rodinia/doku.php?id=gaussian\_elimination}
\showURL{%
\tempurl}


\bibitem[Suite(2018b)]%
        {bfs-rodinia}
\bibfield{author}{\bibinfo{person}{Rodinia~Benchmark Suite}.}
  \bibinfo{year}{2018}\natexlab{b}.
\newblock \bibinfo{booktitle}{\emph{Graph Traversal - Breadth First Search}}.
\newblock
\urldef\tempurl%
\url{https://www.cs.virginia.edu/rodinia/doku.php?id=graph_traversal}
\showURL{%
\tempurl}


\bibitem[Suite(2018c)]%
        {knn-rodinia}
\bibfield{author}{\bibinfo{person}{Rodinia~Benchmark Suite}.}
  \bibinfo{year}{2018}\natexlab{c}.
\newblock \bibinfo{booktitle}{\emph{K-nearest Neighbors}}.
\newblock
\urldef\tempurl%
\url{https://www.cs.virginia.edu/rodinia/doku.php?id=k-nearest_neighbors}
\showURL{%
\tempurl}


\bibitem[Suite(2018d)]%
        {particlefilter-rodinia}
\bibfield{author}{\bibinfo{person}{Rodinia~Benchmark Suite}.}
  \bibinfo{year}{2018}\natexlab{d}.
\newblock \bibinfo{booktitle}{\emph{Particle Filter}}.
\newblock
\urldef\tempurl%
\url{https://www.cs.virginia.edu/rodinia/doku.php?id=particle_filter}
\showURL{%
\tempurl}


\bibitem[Top500(2021)]%
        {top500list}
\bibfield{author}{\bibinfo{person}{Top500}.} \bibinfo{year}{2021}\natexlab{}.
\newblock \bibinfo{title}{Top500 supercomputer sites}.
\newblock
\newblock
\urldef\tempurl%
\url{https://www.top500.org/lists/top500/2022/06/}
\showURL{%
\tempurl}


\bibitem[Tuomi(2002)]%
        {tuomi2002lives}
\bibfield{author}{\bibinfo{person}{Ilkka Tuomi}.}
  \bibinfo{year}{2002}\natexlab{}.
\newblock \showarticletitle{The lives and death of Moore's Law}.
\newblock \bibinfo{journal}{\emph{First Monday}} (\bibinfo{year}{2002}).
\newblock


\bibitem[University(2019)]%
        {Seawulf}
\bibfield{author}{\bibinfo{person}{Stony~Brook University}.}
  \bibinfo{year}{2019}\natexlab{}.
\newblock \bibinfo{title}{SeaWulf, computational cluster at Stony Brook
  University}.
\newblock
\newblock
\urldef\tempurl%
\url{https://it.stonybrook.edu/help/kb/understanding-seawulf}
\showURL{%
Retrieved October 25, 2022 from \tempurl}


\bibitem[Wu et~al\mbox{.}(2022)]%
        {wu2022autotuning}
\bibfield{author}{\bibinfo{person}{Xingfu Wu}, \bibinfo{person}{Michael Kruse},
  \bibinfo{person}{Prasanna Balaprakash}, \bibinfo{person}{Hal Finkel},
  \bibinfo{person}{Paul Hovland}, \bibinfo{person}{Valerie Taylor}, {and}
  \bibinfo{person}{Mary Hall}.} \bibinfo{year}{2022}\natexlab{}.
\newblock \showarticletitle{Autotuning PolyBench Benchmarks with LLVM
  Clang/Polly loop optimization pragmas using Bayesian optimization}.
\newblock \bibinfo{journal}{\emph{Concurrency and Computation: Practice and
  Experience}} \bibinfo{volume}{34}, \bibinfo{number}{20}
  (\bibinfo{year}{2022}), \bibinfo{pages}{e6683}.
\newblock


\end{thebibliography}
\end{document}